# From Atomistic Models to Machine Learning: Predictive Design of Nanocarbons under Extreme Conditions

Xiaoli Yan [1,2], Millicent A. Firestone [1], Murat Keçeli [1], Santanu Chaudhuri [1,2], and Eliu Huerta [1,3,4]*

[1] Argonne National Laboratory, Lemont, Illinois 60439, USA
[2] University of Illinois Chicago, Chicago, Illinois 60607, USA
[3] The University of Chicago, Chicago, Illinois 60637, USA
[4] University of Illinois Urbana-Champaign, Urbana, Illinois 61801, USA

**\*** Correspondence: Eliu Huerta, elihu@anl.gov

**Abstract (300 words maximum)**

The formation of technologically valuable nanocarbon structures under extreme conditions, such as those produced during high-explosive detonations, remains poorly understood but holds significant potential for the development of controlled synthesis pathways. While detonation shockwaves provide the high-pressure, high-temperature environment required for nanodiamond formation, subsequent cooling and decompression dictate whether the diamond phase is preserved or transformed into other nanocarbon structures. Here, we employ GPU-accelerated reactive molecular dynamics (ReaxFF) simulations to investigate the graphitization and structural remodeling of detonation nanodiamond under nonlinear quench and pressure-release trajectories. We further investigate how the initial nanodiamond morphology; cuboctahedral, octahedral, or hexagonal prism influences the resulting transformation products. Evolution of nanostructure, allotrope (via simulated x-ray diffraction), carbon hybridization, and ring statistics are tracked during a two-stage quench from 5000 K and 60 GPa. Rapid cooling combined with slow decompression optimizes cubic diamond retention, whereas slow cooling with rapid pressure release promotes surface-to-core graphitization, producing concentric *sp²*-hybridized layers and hollowed inner shells. Octahedral nanodiamonds evolve into carbon nano-onions, initially forming bucky diamonds that progressively transform into fully sp²-hybridized structures, while hexagonal prisms preferentially form parallel-stacked graphite layers resembling carbon dots. Transient hexagonal diamond (lonsdaleite) emerges as an interfacial phase, suggesting potential reversibility in the shock-induced graphite-to-diamond transformation pathway transformation route. To extend predictive capabilities, we trained machine learning (ML) regressors on over $10^5$ node-hours of molecular dynamics (MD) trajectories. A multilayer perceptron (MLP) model reliably predicts the number of graphitized layers from temperature–pressure trajectories with a coefficient of determination ($R^2$) exceeding 0.90. This high predictive fidelity enables efficient, high-throughput mapping of the synthesis parameter space for optimized graphitization outcomes. Collectively, morphological control combined with optimized quench–decompression conditions promote the selective synthesis of nanocarbon allotropes. This work establishes a data-driven framework for the rational, a priori design of carbon nanomaterials for applications in energy storage, sensing, and biomedicine.

**Keywords**: Detonation nanodiamond remodeling; carbon nano-onions; carbon dots; reactive molecular dynamics, extreme condition synthesis of nanophase carbons, graphitization of nanodiamond, machine learning, multilayer perceptron

**Introduction**

The discovery of new carbon allotropes and the exploration of their unique properties as the basis for novel materials has been a defining aspect of nanoscience. [1, 2] Among these intriguing nanocarbons, several have been identified in detonation soot. [3-5] One of the most notable is nanodiamond (also referred to as detonation diamond), which consists of 3–5 nm spherical particles of cubic diamond. [3] Nanodiamonds have shown tremendous potential in quantum sensing when nitrogen vacancy (NV) centers, or point defects in the diamond lattice, are created. [6] Another remarkable product of explosive detonations is graphitic Onion-Like Carbon particles (OLC), also known as carbon nano-onions. [4] These are quasi-spherical (or polyhedral) nanoparticles with diameters typically less than 10 nm and are composed of multiple concentric fullerene-like layers. Due to their large external surface area and high conductivity, nano-onions are used in electrochemical energy storage applications, such as supercapacitors. [7] Additionally, hollow-core shell graphitic carbon particles, recovered from soot [5] are being investigated for their intrinsic biocompatibility, surface functionalization capabilities, and large interior void space, making them promising candidates for drug delivery systems. [8] Detonation synthesis is a straightforward, single-step process that produces hollow nanoparticles with high yields and fewer steps compared to more traditional methods, which typically require complex templating and etching to create large internal voids.

Although a variety of novel nanophase carbons have been identified in detonation soot, little is known about the chemical reactions occurring behind the detonation front or how these nanocarbon structures undergo remodeling (i.e., by, for example, graphitization) during pressure release and cooling. Variations in detonation conditions (e.g., chemical composition, peak temperature/pressure, atmosphere, and cooling /pressure release rates) can lead to reproducible differences in nanocarbon products. [5, 9, 10] For example, it is well established that detonations using Composition B and an ice quench step reliably yield nanodiamonds and are routinely employed for their commercial production. [11] In contrast, Composition B detonations carried out in a closed chamber under an ambient air atmosphere excluding the ice quench step result in quasi-spherical (polyhedral) hollow-core particles (with an average core radius of 22.9 ± 3.8 nm) and lamellar shells (10.2 ± 5.4 nm thick) composed of $sp^2$-hybridized carbon, indicating significant graphitization. [5] Similarly, nanoparticles produced in detonations under an inert argon atmosphere were highly anisotropic, discoidal, and lacked internal voids, with an average circumferential radius of 35.1 ± 0.5 nm and a thickness of 4.3 ± 0.7 nm. [5] These particles were also composed of $sp^2$-hybridized carbon but exhibited primarily planar carbon hexagonal structures, in contrast to the non-planar (curved) carbon frameworks (mix of hexagonal and pentagonal rings) found in particles produced in air. This clearly suggests that the detonation atmosphere and/or cooling medium play crucial roles in determining the resulting nanoparticle structures.

A lot of attention has focused on the study of nanodiamond graphitization into onion-like structures. [12] Recent studies examined the formation of OLCs by graphitization of nanodiamond by thermal annealing at temperatures up to 1100 °C in different atmospheric conditions (vacuum, inert gas). These studies found that within the 800–1100 °C temperature range, precise control over the amount and crystallinity of $sp^2$-hybridized carbon at the periphery of the nanodiamonds could be achieved. Additionally, a higher fraction of $sp^2$ carbon was formed in a vacuum environment compared to an argon atmosphere. [13] While this prior work provides insights into graphitization-induced remodeling of nanodiamonds, further research is needed to enable the reliable production of the full spectrum of novel nanophase carbons formed under extreme conditions. Moreover, advancing our understanding of the chemistry and physics of carbon under extreme temperatures (500–5000 K) and pressures (0.5–200 GPa) is essential not only for materials synthesis under extreme conditions but also important for enhancing our fundamental knowledge of explosive performance, astrochemistry, and geochemistry (e.g., Earth's mantle and core). [14-17] Due to the experimental difficulties and high costs associated with studies under extreme conditions, the number of experiments is severely limited, making detailed investigations challenging. As an alternative, computational models simulating these processes may provide a viable means of gaining additional insights and generating more data to support machine learning (ML) and artificial intelligence (AI). In this study, we present a computational method for simulating the graphitization and remodeling of detonation diamonds (nanodiamonds) as a function of cooling and pressure-release rates following detonation.

Over the past 30 years, numerous computational studies have attempted to model the surface graphitization process of diamond materials. [18] These studies have employed various computational methods, ranging from first-principles calculations to Density Functional Tight Binding (DFTB) methods,[19, 20] and Molecular Dynamics (MD) simulations using different force fields. Barnard et al. used first-principles methods to study the surface graphitization of finite-sized (323 atoms) nanodiamond on the {111}, {110}, and {100} crystal planes. [21, 22] Barnard et al. achieved 87% conversion of nanodiamond to carbon nano-onions (OLCs) using DFTB with 1798 atoms. [23] Others have used MD simulations with a Reactive Empirical Bond Order (REBO) force field to anneal at elevated temperatures (ca. 2000 K) spherical nanodiamond into OLC structures achieving conversion of up to 87%. [24-26] Adiga et al. achieved conversion to 97% at 2500 K. [27] In addition, studies carried out by others have successfully produced OLCs on simulations with a larger number of atoms (5000-7000) using advanced force fields such as Long Range Carbon Bond Order Potential (LCBOP) and ReaxFF. [28-30] The achievement of 99% conversion of 55272 atoms of octahedral nanodiamond to nano-onions at 3000 K was achieved by Dai et al. [31, 32] Collectively, these studies on thermal graphitization of diamond are important and foundational to our work providing, most notably, insights on the formation of ordered graphitic carbon.

Beyond the graphitization of nanodiamond, the work of Powles et al. [33] investigated the structural evolution of disordered carbon (i.e., amorphous carbon precursors) under high-temperature annealing (4000 K) using an environment-dependent force field. Their study demonstrated that ordered $sp^2$ carbon phases spontaneously emerge at elevated temperatures through a thermodynamically driven process. These phases self-assemble into highly ordered $sp^2$-bonded networks, with the degree of ordering influenced by both external factors (such as the initial geometry) and internal features (such as voids and free surfaces) of the amorphous precursors, as the ordering is initiated at the surfaces. Amorphous polymer carbon films have also been simulated, and it was found that increasing the annealing temperature enhances the degree of graphitization within the carbon matrix. This manifests as reconstructed amorphous carbon fragments resulting from $sp^3$ to $sp^2$ hybridization transformations. [34]

While prior work has examined how temperature influences the graphitization of nanodiamond and the ordering of disordered *sp²* carbon, the role of pressure in the formation and remodeling of carbon structures remains largely unexplored. Bidault et al. modeled the formation of carbon nanostructures, including diamond, by simulating high-pressure pyrolysis using dynamic MD simulations with a ReaxFF force field. [35] Alternatively, Enriquez et al. developed a ML force field to study the surface graphitization of bulk diamond structures in vacuum and reported the diamond {111} surface is more susceptible to thermal degradation. [36] Ostroumova et al. conducted a reactive MD study on the ultrafast cooling of pure carbon gas from 6000 to 2500 K. The simulations revealed that the carbon gas initially transforms into a predominantly *sp*-hybridized, low-density polymer gel, which then transitions into a dense *sp²*-hybridized liquid particle before ultimately evolving into a fullerene-like structure. The graphitization of the carbon nanoparticle occurs abruptly within a narrow temperature range (T = 4100–4150 K), beginning at the particle surface and forming fullerene-like shells sequentially until the entire nanoparticle adopts an onion-like morphology. [37] Building on this prior work, Orekhov et al. evaluated three different models for reactive MD simulations of OLC formation from gas-phase carbon and extended the pressure range studied to 0.1 GPa. Pressure was introduced by conducting high-temperature simulations in a constant-volume box. [38]

In this work, we demonstrate how reactive molecular dynamics (MD) simulations can be leveraged to elucidate the influence of nonlinear cooling and pressure-release protocols on the graphitization of detonation nanodiamonds, with the aim of reliably producing distinct nanocarbon phases. Building on this mechanistic insight, we develop a predictive framework by training several machine learning (ML) regression models on over $10^5$ node-hours of MD trajectory data.

## Methods

**Reactive Molecular Dynamics (MD) Simulations.**

The formation of detonation nanodiamonds under high-pressure, high-temperature (HPHT) conditions is well established. [39, 40] In this study, given a peak initial temperature (5000 K) and pressure (60 GPa), the simulations assume that nanodiamond has already formed. The focus of the computational modeling is on the subsequent transformation and particle remodeling that occur during cooling and pressure release. Reactive molecular dynamics simulations are performed using LAMMPS to model these processes. [41] Initial nanodiamond morphologies are generated from a cubic diamond supercell, partitioned using plane equations. Pressure is applied in the simulations by introducing argon atoms as filler. The simulations begin from an equilibrated state at 5000 K and 60 GPa, followed by simultaneous, non-linear decreases in temperature and pressure until reaching a final state of 300 K and 1 atm.

*Reactive Force Field*. The ReaxFF (Reactive Force Field) model, developed by van Duin et al. for simulating hydrocarbon systems, is an empirical, bond-order-based potential designed to model reactive molecular environments. [42] Unlike traditional force fields that assume fixed bond topologies, ReaxFF enables the dynamic formation and dissociation of chemical bonds during simulations, making it particularly suitable for studying chemical reactions under a wide range of conditions. ReaxFF has been extended to a variety of materials, including carbon-based nanostructures such as graphene, nanotubes, and fullerenes. It provides a quantitative framework for modeling reactions in both molecular and crystalline systems. By capturing both covalent and non-covalent interactions and adapting to changes in atomic environments, ReaxFF offers a powerful tool for exploring chemical reactivity and material behavior across diverse applications. The potential energy of a system in ReaxFF is evaluated using the following equation:

$$E_{system} = E_{bond} + E_{over} + E_{under} + E_{lp} + E_{val} + E_{tor} + E_{vdW} + E_{Coulomb} + E_{trip} \qquad (1)$$

where the $E_{bond}$ is the bond interaction energy, $E_{over}$ is the over-coordination energy penalty, $E_{under}$ is the under-coordination energy penalty, $E_{lp}$ is the lone-pair electron energy, $E_{val}$ is the valence angle energy, $E_{tor}$ is the torsion angle energy, $E_{vdW}$ is the van der Waals energy, $E_{Coulomb}$ is the Coulomb energy, and $E_{trip}$ is the triple bond stabilization energy. The ReaxFF and $Q_{eq}$ charge equilibration packages from LAMMPS is used to conduct ReaxFF simulations in this work. [43]

Based on a few previous computational studies in the diamond/graphite/fullerene area [44-47] the carbon ReaxFF parameters from Damirchi, et al. [48], an improved version of the ReaxFF-2013, [49] are used in this work. For the argon atoms, the ReaxFF parameter is chosen from Yoon et al.'s work. [50] Recent comparative studies of carbon interatomic potentials have shown that different force fields exhibit markedly different capabilities in describing diamond–graphite energetics and sp²↔sp³ phase transformations. Short-range bond-order potentials such as REBO and BOP reproduce diamond lattice constants but lack adequate long-range interactions, leading to limitations in modeling pressure-driven phase stability. AIREBO introduces long-range Lennard–Jones interactions but can over-stiffen interlayer response and distort elastic and transformation energetics. In contrast, LCBOP provides highly accurate diamond–graphite energetics and transition pathways, at the cost of substantially higher computational expense. ReaxFF occupies an intermediate position: it preserves the correct ordering of graphite as the stable phase at low pressure and diamond at high pressure, reproduces reasonable cohesive energies and lattice constants for sp² and sp³ carbon, and enables reactive bond breaking and formation under extreme conditions. While ReaxFF underestimates the absolute diamond–graphite energy separation relative to LCBOP, it captures the correct phase tendencies and pressure dependence, which is the critical requirement for the mechanistic trends explored here.

An additional motivation for using ReaxFF in this work is its compatibility with GPU-accelerated molecular dynamics in LAMMPS, which enables simulations of $10^4$–$10^5$ atom systems over long time scales and across large ensembles of thermodynamic trajectories. This capability is essential for systematically exploring nonlinear cooling and pressure-release pathways and for generating the large, consistent datasets required for downstream machine-learning analysis.

*Initial Geometry of Nanodiamonds.* Based on experimental observations, low index crystal planes, i.e. {111}, {110}, {100} planes, are the most common ones that occur on the nanodiamond surfaces. [51] In a computational model, derived by Barnard et al. to predict the thermodynamically favorable morphology, only low index crystal planes are considered for simplicity. [21] We do recognize the computational studies conducted on the high index crystal planes, such as {331} and {221}, yielded lower surface energy values compared to those of the low index crystal planes. [52, 53] In fact, in natural microdiamonds (diameter 0.1-1.0 mm), high index crystal planes are observed experimentally via scanning electron microscopy (SEM). [54] However, due to the scarcity of supporting experimental observations in nanodiamond, we limited the scope of our computational study to low index crystal planes: {111}, {110}, and {100}.

Initial simulations were performed on nanodiamonds with cubic morphology (exposing exclusively {100} facets) and rhombic dodecahedral morphology (exposing exclusively {110} facets), with no evidence of stacked graphite-like structures observed in either case. To facilitate the unidirectional growth of an oriented graphite structure, the initial nanodiamond configuration must contain exactly one pair of parallel {111} crystal planes. This unique pair of {111} planes is expected to govern the formation of surface graphitized structures, while all other surfaces should not belong to the {111} family. For reasons of symmetry and simplicity, a set of six {110} planes is selected to define the remaining facets of the designed initial geometry.

The unit cell structure mp-66 from the Materials Project is used to generate the initial nanodiamond structure. [52] This unit cell is replicated in the $x$, $y$, and $z$ directions to form a cubic supercell centered at the origin (0,0,0). To create a cubic supercell with edge length $L$, the boundaries of the cube are described by the plane equations of the {100} crystal planes, which can be written as:

$$x = \pm \frac{L}{2}, y = \pm \frac{L}{2}, z = \pm \frac{L}{2} \qquad (2)$$

A cuboctahedral nanodiamond is generated by initially constructing a cubic supercell. Next, a set of eight equivalent {111} crystal planes are defined, which intersect the supercell cube at the midpoints of all cube edges. These points are given by the following coordinates:

$$\left(0, \pm\frac{L}{2}, \pm\frac{L}{2}\right), \left(\pm\frac{L}{2}, 0, \pm\frac{L}{2}\right), \left(\pm\frac{L}{2}, \pm\frac{L}{2}, 0\right) \tag{3}$$

The corresponding plane equations that describe these {111} crystal planes are:

$$\pm x \pm y \pm z = L \tag{4}$$

The created cuboctahedron nanodiamond is shown in **Fig. 1(A).** The structure has 21,830 atoms and radius of gyration, $R_g$, of 2.43 nm.

Alternatively, an octahedral nanodiamond can be created by starting with a cubic supercell with edge length $L$. A set of eight equivalent {111} crystal planes intersects the supercell at the following points:

$$\left(0, 0, \pm\frac{L}{2}\right), \left(0, \pm\frac{L}{2}, 0\right), \left(\pm\frac{L}{2}, 0, 0\right) \tag{5}$$

The corresponding plane equations for these {111} crystal planes are:

$$\pm x \pm y \pm z = \frac{L}{2} \tag{6}$$

The resulting octahedral nanodiamond is shown in **Fig. 1E**. The structure contains 10,660 atoms and has a $R_g$ of 1.89 nm.

To create a hexagonal prism nanodiamond, the cubic supercell is truncated by {111} and {110} crystal planes. The {111} planes intersect the supercell at the following points:

$$\left(-\frac{L}{2}, \pm\frac{L}{2}, \frac{L}{2}\right), \left(\frac{L}{2}, \pm\frac{L}{2}, 0\right), \left(\pm\frac{L}{2}, \frac{L}{2}, -\frac{L}{2}\right) \tag{7}$$

while the {110} planes intersect the supercell at:

$$\left(0, \pm\frac{L}{2}, \pm\frac{L}{2}\right), \left(\pm\frac{L}{2}, 0, \pm\frac{L}{2}\right), \left(\pm\frac{L}{2}, \pm\frac{L}{2}, 0\right) \tag{8}$$

The plane equations for the {111} crystal planes can be written as:

$$x + y + z = \pm\frac{L}{2} \tag{9}$$

while the plane equations for the {110} crystal planes are:

$$x - y = \pm\frac{L}{2}, y - z = \pm\frac{L}{2}, x - z = \pm\frac{L}{2} \tag{10}$$

The created hexagonal prism nanodiamond is shown in **Fig. 1(I)**. The structure has 12,315 atoms and a $R_g$ of 1.98 nm.

Carbon atoms located outside these boundaries are removed, and the remaining atoms constitute the desired nanodiamond shape, which in this case is a cuboctahedron. The crystal manipulation is performed with the assistance of the Pymatgen library. [54] These plane equations are used not only for removing excess carbon atoms but also for adding argon atoms. The region inside the boundaries defined by the plane equations corresponds to the nanodiamond with the desired crystal structure, while the region outside is filled with argon atoms. The argon atoms act as the medium for applying external isostatic pressure to the nanodiamond during the simulations.

*Simulation Configuration*. The simulation box is defined to be orthogonal with periodic boundary conditions in the $(x, y, z)$ directions. Argon atoms are introduced as an inert pressure-transmitting medium to impose hydrostatic loading on the nanodiamond. Noble gases, and argon in particular, are widely used as pressure-transmitting media in diamond anvil cell experiments because they minimize deviatoric stresses and provide near-hydrostatic conditions up to tens of GPa. [55] The elastic properties and equation of state of solid argon have been experimentally characterized to pressures exceeding 60–70 GPa, demonstrating that compressed argon remains a mechanically coherent medium capable of transmitting isotropic stress. [56, 57] Shock-compression and equation-of-state studies further show that argon remains thermodynamically stable and well described at pressures far exceeding those considered here, supporting its use as a confining medium under extreme conditions. [58] While static experiments do not access temperatures as high as 5000 K, in MD simulations an equilibrated argon bath provides an effective mechanical approximation to hydrostatic confinement provided that the pressure tensor remains isotropic and no Ar–C bonding occurs (see below). Both conditions were explicitly verified by monitoring the virial stress tensor components and Ar–C coordination throughout representative trajectories.

The Ar–C interactions are described using ReaxFF parameters for argon that are purely nonreactive, ensuring that argon does not form chemical bonds with carbon and serves only as a mechanical confining medium. This inertness assumption is standard and is supported by the absence of any observed Ar–C bonding or charge transfer during equilibration and quench–release trajectories. Hydrostatic pressure is maintained by evolving the full system (nanodiamond plus argon) in an isotropic NPT ensemble using a Nosé–Hoover thermostat and barostat, which allows the simulation cell to fluctuate uniformly in all directions. During cooling and pressure release, the target temperature and pressure are continuously ramped according to the prescribed nonlinear trajectories, while the Nosé–Hoover chains ensure proper sampling of the thermodynamic ensemble and suppress spurious pressure anisotropy. This approach provides an accurate representation of hydrostatic confinement and release throughout the quench process.

An isotropic isothermal-isobaric ensemble (NPT) is applied such that the system starts with ($P_0$, $T_0$) and ends with ($P_f$, $T_f$). The simulation is divided into two stages with different combinations of quench rates and pressure-release rates. Before the quench simulation starts, each nanodiamond/argon system is placed under 60 GPa and annealed from 1 K to 5000 K in 2,000,000 steps. Then the system is equilibrated at 5000 K and 60 GPa for another 1,000,000 steps. Both processes employ an isotropic NPT ensemble and a time step of 0.01 fs. While ReaxFF inherently requires smaller time steps than non-reactive MD, even finer time steps are necessary for ReaxFF simulations of highly excited or reactive carbon systems, where large forces, steep potential energy surfaces, and rapid bond-order changes can otherwise lead to numerical instabilities and poor energy conservation. For example, Hashemi et al., employed a 0.01 fs time step in ReaxFF simulations of energetic hydrocarbon fragmentation to maintain stability over nanosecond trajectories. [59] In the present work, the 0.01 fs time step was found to ensure stable integration and consistent energy behavior under the combined conditions of high temperature (~5000 K), high pressure (~60 GPa), and rapid quench dynamics; larger time steps led to noticeable integration noise and were therefore not used. The purpose of this procedure is to ensure that the nanodiamond remains well-equilibrated at conditions of peak temperature and peak pressure while preserving the cubic diamond structure. After the equilibration, the nanodiamond system undergoes a two-stage continuous cooling and pressure loss process. In the first stage, the temperature $T_0$ is lowered to $T_{mid}$ linearly, and the pressure $P_0$ is lowered to $P_{mid}$ linearly. In the second stage, the temperature $T_{mid}$ is lowered to $T_f$ linearly, and the pressure $P_{mid}$ is lowered to $P_f$ linearly. The temperature and pressure change rates are different in the two stages, and their values depend on the choice of $(T_{mid}, P_{mid})$. Previous atomistic studies of nanocarbon formation have typically employed vacuum conditions with single-stage linear cooling, omitting the effects of pressure release during expansion. Here, we extend this framework by explicitly including pressure evolution and a two-stage, piecewise-linear pressure–temperature trajectory with a defined intermediate state, providing a more realistic abstraction of detonation thermodynamics. The intermediate ($P_{mid}$, $T_{mid}$) captures the transition between extreme detonation conditions and expansion-driven quenching and is shown to play a dominant role in determining final carbon morphology. A variable time-step strategy is employed to ensure numerical stability at extreme conditions while reducing computational cost by approximately 50%, enabling efficient exploration of pathway-dependent effects. The combination of simulation quench and pressure-release rates used in this study are summarized in **Table 1**.

    We note that the cooling and pressure-release rates accessible to classical MD simulations (typically on picosecond timescales) are necessarily much faster than those realized in experimental detonations, where pressure release and thermal quenching occur over microsecond to millisecond timescales depending on explosive formulation and confinement conditions. This timescale disparity is intrinsic to atomistic MD and is common to prior simulations of nanodiamond graphitization, carbon nano-onion formation, and ultrafast carbon condensation under extreme conditions. Consequently, the present simulations are not intended to reproduce absolute experimental kinetics or transition rates. Instead, our conclusions focus on relative trends, namely, how variations in cooling versus pressure-release pathways and initial nanodiamond morphology bias competing sp³-to-sp² transformation mechanisms within a consistent dynamical framework. These relative dependencies are robust to uniform rescaling of rates and provide mechanistic insight into experimentally observed sensitivities of detonation nanocarbon products to post-detonation thermodynamic histories.

**Table 1**. Summary of the combinations of simulation quench and pressure-release rates used in this work.

| Temperature (K) | | | Quench Rate (K/ps) | |
|---|---|---|---|---|
| $T_0$ (peak T) | $T_{mid}$ | $T_f$ | $\Delta T_1/\Delta t$ | $\Delta T_2/\Delta t$ |
| 5000 | 4000 | 300 | 10 | 37 |
| 5000 | 3000 | 300 | 20 | 27 |
| 5000 | 2000 | 300 | 30 | 17 |
| Pressure (GPa) | | | Release Rate (GPa/ps) | |
| $P_0$ (peak P) | $P_{mid}$ | $P_f$ | $\Delta P_1/\Delta t$ | $\Delta P_2/\Delta t$ |
| 60 | 15 | $1.01\times10^{-4}$ | 0.450 | 0.150 |
| 60 | 6.0 | $1.01\times10^{-4}$ | 0.550 | 0.050 |
| 60 | 2.5 | $1.01\times10^{-4}$ | 0.575 | 0.025 |
| 60 | 2.0 | $1.01\times10^{-4}$ | 0.580 | 0.020 |
| 60 | 1.5 | $1.01\times10^{-4}$ | 0.585 | 0.015 |
| 60 | 1.0 | $1.01\times10^{-4}$ | 0.590 | 0.010 |
| 60 | 0.5 | $1.01\times10^{-4}$ | 0.595 | 0.050 |

*Polyhedral Template Matching (PTM) Analysis*. The carbon nanoparticle is composed of carbon atoms exhibiting a variety of hybridization states and local crystal structures. The Polyhedral Template Matching (PTM) algorithm is employed to systematically identify the local atomic environments of carbon atoms at each frame of the simulation trajectory. [54] The PTM package within the LAMMPS software is used to provide real-time atom typing information. In this study, each carbon atom is classified as one of the following local crystal structure types: cubic diamond (CUB), hexagonal diamond (i.e., lonsdaleite, HEX), or the $sp^2$ hybridized state (SP2). To quantify the degree of graphitization, we define the number of $sp^2$ layers using a geometry-based radial segmentation algorithm that is robust to curvature and partial disorder.

At each saved MD frame, carbon atoms are first classified as $sp^2$ using the PTM analysis. The particle center is defined by the center of mass of all carbon atoms, and the radial distance r of each $sp^2$ atom from this center is computed. A one-dimensional radial number-density profile of $sp^2$ atoms is then constructed by binning r with a fixed bin width. Distinct graphitic shells manifest as pronounced peaks in this radial density profile, corresponding to concentric $sp^2$-rich layers, even for curved or non-spherical morphologies. The number of $sp^2$ layers is defined as the number of statistically significant density maxima, identified using a peak-finding criterion based on local prominence and minimum peak-to-peak separation. This procedure naturally accommodates curved, buckled, and partially disordered shells and avoids reliance on planar geometry assumptions.

*Ring Analysis*. For each frame of the simulation trajectory, the SP2 carbon atoms identified by the PTM algorithm are further analyzed for number of N-size (N $\in$ [4, 8]) rings. The SP2-labelled carbon atoms' Cartesian coordinates are used to identify the covalent bonds via the k-d tree [60, 61] and pairwise distance method implemented in SciPy [62]. The bonding information are fed to the NetworkX [63] to generate a molecular graph of all the SP2-labeled carbon atoms. The Johnson-Gupta-Suzumura algorithm [64-68] implemented in NetworkX is applied on the molecular graph to extract the number of N-size rings. A N-size ring in the chemistry context can be viewed as a simple cycle in the mathematics context. A molecular graph with all *sp*2 hybridized atoms is constructed with NetworkX, then the simple cycles of length N (N $\in$ [4, 8]) are counted. The resulting integer layer count, extracted consistently across all trajectories and time frames using this same algorithm, is used as the target variable for machine-learning regression. Because the method is based on radial density peaks rather than idealized crystallinity or perfect registry, it remains well defined throughout the quench–release process, including transient bucky-diamond states and partially graphitized structures.

*Simulated X-ray Diffraction Pattern.* X-ray diffraction patterns are generated using the LAMMPS "compute xrd" from the Diffraction package [69]. Cu-K$\alpha$ X-ray with wavelength 1.54 Å is used as the X-ray source.

The $2\theta$ angles are considered from 10° to 100°. The x-ray diffraction intensity $I$ at angle $\theta$ can be expressed as:

$$I = L_p(\theta)\frac{|F|^2}{N} \quad (11)$$

$F$ is the structure factor:

$$F(\vec{k}) = \sum_{j=1}^{N} f_j(\theta)\, e^{2\pi i \vec{k}\cdot\vec{r}_j} \quad (12)$$

where $rj$ is the j-th atom's position in the real space, $fj$ is the j-th atom's scattering factor.

$Lp$ is the Lorentz polarization factor:

$$L_p(\theta) = \frac{1+\cos^2(2\theta)}{\cos(\theta)\sin^2(\theta)}, \quad (13)$$

The scattering angle $\theta$ and the reciprocal lattice vector $\vec{k}$ satisfy Bragg's law [70]:

$$\frac{\sin(\theta)}{\lambda} = \frac{|\vec{k}|}{2} \quad (14)$$

To convert the domain variable $2\theta$ to scattering vector's magnitude $|\vec{q}|$, the following equation was used:

$$\frac{4\pi}{\lambda}\sin(\theta) = |\vec{q}| \quad (15)$$

This simulated x-ray diffraction pattern is generated for key frames of the simulation. X-ray diffraction patterns are also generated using the LAMMPS compute xrd command on selected frames during the quench process. X-ray diffraction patterns for bulk diamond and bulk graphite are also generated as baseline values.

**Machine Learning (ML) Prediction.**

To investigate the nonlinear relationship between thermodynamic conditions and structural transitions in nanodiamond, a suite of machine learning (ML) models was trained on curated MD datasets. The objective was to develop predictive models capable of estimating the degree of graphitization from temperature and pressure variations encountered during quenching and pressure-release protocols. A range of regression algorithms, including B-spline, gradient boosting, random forest (RF), and multilayer perceptron (MLP) along with their associated hyperparameter configurations, are summarized in **Table 2**. To mitigate overfitting, an 80:20 train–test data split was employed. All models were implemented using the scikit-learn library version 1.6.1. [71] Input data were obtained from our MD simulations of nanodiamond graphitization. Specifically, the features correspond to normalized progress in temperature and pressure changes across successive stages of the simulation, expressed as percentages of the total variation. The output variable is the number of graphitized carbon layers on the nanodiamond surface. For an octahedral nanodiamond with nine potential graphitizable layers, we illustrate the scheme with a representative case: four layers undergo graphitization under a quench and release pathway characterized by intermediate conditions of $P_{mid}$ = 6 GPa and $T_{mid}$ = 2000 K. The total thermal excursion is $\Delta T_{total}$ = 4700 K, partitioned into $\Delta T_1$ = 3000 K (63.83%) and $\Delta T_2$ = 1700 K (36.17%). The total pressure drop is $\Delta P_{total}$ = 60 GPa, partitioned into $\Delta P_1$ = 54 GPa (95.83%) and $\Delta P_2$ = 6 GPa (4.17%). Accordingly, the input vector for the ML model is represented as a 4D tensor: X = [63.83, 36.17, 95.83, 4.17], where the first two entries encode fractional contributions to temperature change, and the latter two encode fractional contributions to pressure loss. Due to the size of our simulation dataset, we curated the training-test dataset split to ensure statistical robustness while preserving the physical integrity of the underlying simulation trajectories. The ML models developed here are trained on simulations with fixed particle size and bounded thermodynamic conditions, reflecting the computational cost of generating reactive MD trajectories. Within this domain, the models exhibit robust predictive performance on unseen trajectories and capture physically interpretable dependencies on cooling and pressure-release histories, demonstrating generalization across dynamical pathways rather than overfitting to individual simulations. We emphasize that the ML models are not intended to extrapolate beyond the sampled size or P–T space; instead, they provide a quantitatively reliable mapping between thermodynamic history and resulting sp² layering within the defined domain. Extension to additional sizes, compositions, or pressure–temperature ranges would require corresponding expansion of the training data. We have released the scientific software to reproduce these results, including methodological information and implementation details for each ML model, and interactive visualizations in the GitHub repository: https://github.com/williamyxl/nanodiamond.

Table 2. Summary of key hyperparameters for the evaluated ML models

| Hyperparameters | B-spline | Gradient Boosting | Random Forest | Multilayer Perception |
|---|---|---|---|---|
| Depth / Layers | 8 | 8 | 8 | 8 |
| Unit per Layer | 8 Knots | 100 estimators | 100 estimators | 8 neurons |
| Learning Rate | N/A | 0.1 | N/A | 0.001 |

**Results & Discussion**

Herein, we present the application of dynamic (reactive) molecular dynamics (MD) simulations using the GPU-accelerated ReaxFF (Reactive Force Field) [42], implemented in the LAMMPS package [41], to investigate the effects of temperature cooling and pressure-release rates on the graphitization-induced remodeling of detonation-formed nanodiamond. Unlike prior studies, our work explicitly considers both pressure conditions and variations in nanodiamond surface structure. Most importantly, this study explores the interplay between the initial nanodiamond morphology and post-detonation processing parameters, specifically, temperature cooling and pressure-release rates and how this combination of factors can be leveraged to engineer unique nanophase carbon materials.

In the simulations, 10,660 to 21,830 carbon atoms are included, and hydrostatic pressure is applied by introducing up to 110,000 argon atoms into the simulation box. All simulations assume an initial nanodiamond particle with radius of gyration $R_g$ = 1.89 to 2.43 nm (radius ∼ 2.00 - 3.84 nm) with varying morphologies. These different morphologies are generated by selectively terminating surfaces of the face-centered cubic diamond crystal structure, resulting in polyhedral particles with a single dominant crystal face, such as {100} for cubic, {111} for octahedral (**Fig. 1E**), and {110} for dodecahedral shapes. More complex surface morphologies are also explored, including combinations of crystal faces: {110} and {100} to form cuboctahedra (**Fig. 1A**), and {111} and {110} to create a hexagonal prism (**Fig. 1I**). The hexagonal prism geometry is introduced here not as a representative detonation nanodiamond morphology, but as a simplified, anisotropic model system intended to explore potential facet- and orientation-dependent shear responses. Prior studies suggest that certain diamond facets and stacking directions may be more susceptible to shear-driven reconstruction; the prism geometry serves to accentuate these directional tendencies within a controlled framework. The crystal structures are derived from a cubic diamond supercell, with particle partitioning achieved using plane equations. Each nanodiamond surrounded by argon atoms is first annealed using 1,000,000 steps of NPT ensemble with temperature ramped from 1 K to 5000 K and pressure kept at 60 GPa. Then, the system is maintained at the peak temperature ($T_0$ = 5000 K) and peak pressure ($P_0$ = 60

GPa) for another 1,000,000 steps. The time step size used here is 0.1 fs, so the annealing process takes 100 ps, and the following stage takes 100 ps. After equilibration, the nanodiamond system was subjected to a two-stage continuous cooling and pressure loss process. The cooling and pressure loss trajectories are superimposed on the carbon phase diagram (**Fig. 2**). All trajectories begin within the diamond region of the phase diagram and traverse and end in the graphite region. [72]

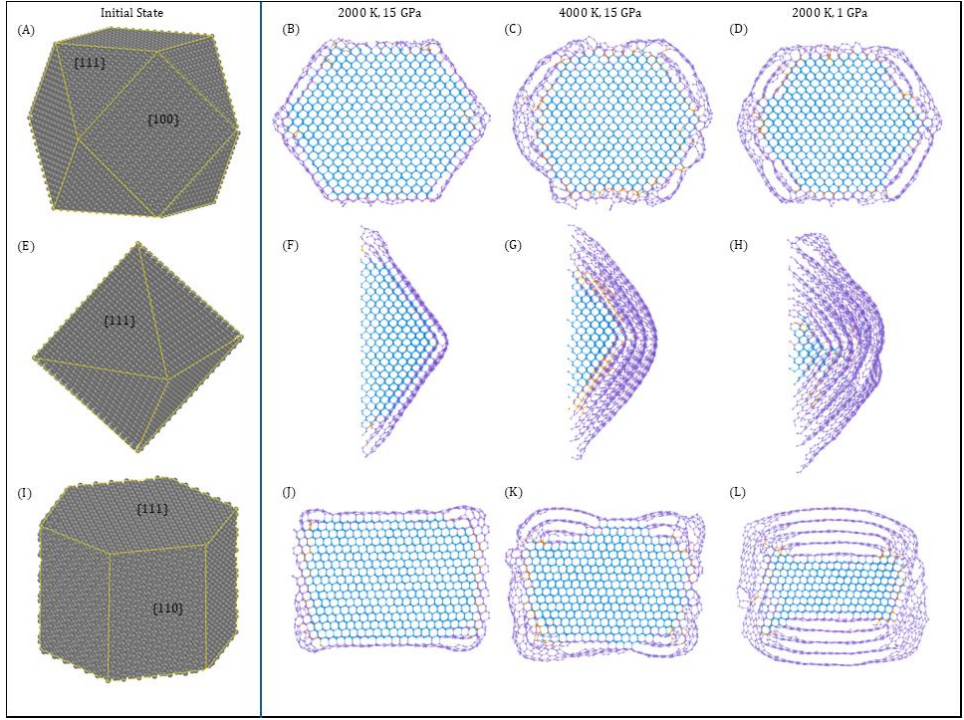

**Figure. 1**: Nanodiamond initial geometries of (A) cuboctahedron with {111} and {100} planes. (E) octahedron with {111} planes. (I) hexagonal prism with {111} and {110} planes (Yellow lines are visual guides). Center slice with a thickness of 10 Å for each final state of the cuboctahedron at: (B) $T_{mid}$ = 2000 K and $P_{mid}$ = 15 GPa; (C) $T_{mid}$ = 4000 K and $P_{mid}$ = 15 GPa; (D) $T_{mid}$ = 2000K and $P_{mid}$ = 1 GPa. Center slice with a thickness of 10 Å for each final state of the octahedron at: (F) $T_{mid}$ = 2000 K and $P_{mid}$ = 15 GPa; (G) $T_{mid}$ = 4000 K and $P_{mid}$ = 15 GPa; (H) $T_{mid}$ = 2000K and $P_{mid}$ = 1 GPa. Center slice with a thickness of 10 Å for each final state of the hexagonal prism at: (J) $T_{mid}$ = 2000 K and $P_{mid}$ = 15 GPa; (K) $T_{mid}$ = 4000 K and $P_{mid}$ = 15 GPa; (L) $T_{mid}$ = 2000K and $P_{mid}$ = 1 GPa.

**Retention of Nanodiamonds**. To validate the simulation approach, we begin by confirming the model produces the expected experimentally established result that rapid cooling upon pressure release preserves nanodiamond (i.e., inhibits graphitization). As shown in **Fig. 2** (purple dotted line top trajectory) the simulation is performed using an initial rapid temperature drop ($\Delta T_l / \Delta t$ = 30 K/ps) under slow pressure release ($\Delta P_1/\Delta t$ = 0.45 GPa/ps) from the peak T and P until a midpoint is reached at $T_{mid}$ = 2000 K and $P_{mid}$ = 15 GPa. This is followed by a second stage of cooling and pressure loss continued at slower rates ($\Delta T_2/\Delta t$ = 17 K/ps; $\Delta P_2/\Delta t$ = 0.15 GPa/ps) until the final conditions are attained ($T_f$ = 300 K and $P_f$ = 1 ×10$^{-4}$ GPa (or 1 atm)). The effect of post-detonation expansion under conditions of rapid temperature reduction and slower pressure release on cuboctahedral nanodiamond is shown in **Fig. 1B-D**. A cross-section of the final product, obtained by slicing the particle through its center, reveals that the nanodiamond remains largely unaffected, with most of the structure preserved as $sp^3$ cubic diamond (blue), and only a single surface layer of graphitic $sp^2$ carbon (purple) (**Fig. 1B**) formed.

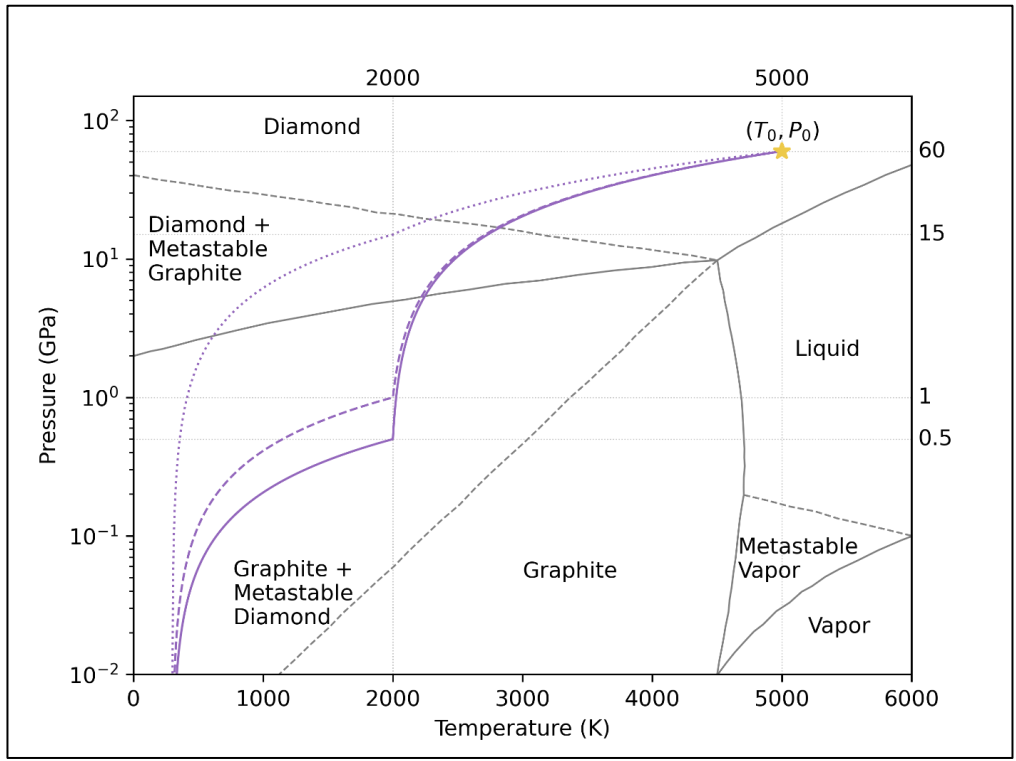

**Figure 2**: Visualization of the nonlinear T-P trajectories superimposed on the carbon phase diagram. Purple solid curve indicates the $T_{mid}$ = 2000 K and $P_{mid}$ = 0.5 GPa. Purple dashed curve indicates the $T_{mid}$ = 2000 K and $P_{mid}$ = 1.0 GPa. Purple dotted curve indicates the $T_{mid}$ = 2000 K and $P_{mid}$ = 15 GPa. The yellow star indicates the starting condition with $T_0$ = 5000 K, $P_0$ = 60 GPa. Pressure (Y) axis is in logarithmic scale. The phase boundaries are extracted from a published technical report. [72]

The impact of reducing the cooling rate to 10 K/ps during the initial cooling phase, down to $T_{mid}$ = 4000 K along with a comparable pressure release ($\Delta P_l/\Delta t$ =0.45 GPa/ps to $P_{mid}$=15 GPa). As expected, the slower cooling rate promotes the growth of additional graphitic surface layers. Furthermore, the graphitic layers appear disordered and partially dissociated from the surface. Dissociation of carbon atoms from the nanodiamond surface has been previously simulated in studies examining the relative stability of different crystal faces in oxygen-containing environments. This surface etching or carbon atom dissociation can lead to fundamentally different particle shapes, depending on the extent and crystallographic orientation of the degradation. [73] The increase in $sp^2$ carbon with slower cooling is consistent with previous studies, which demonstrated that the yield of detonation nanodiamond decreases with an increase in the heat capacity of the detonation chamber medium (and thus slower cooling). [74] Further noted are the core of the particle (**Fig. 1C**) remains composed of $sp^3$ cubic diamond and an interfacial region of $sp^3$ hexagonal carbon (orange) is more clearly observed (as compared to **Fig. 1B**). While prior work has explored the shock-induced formation of metastable lonsdaleite from highly ordered graphite, [75] our findings reveal its formation not during shock compression, but rather as an intermediate transition layer during the graphitization of cubic diamond. Unlike the well-established transformation of graphite into cubic diamond via a hexagonal diamond intermediate under shock conditions, the simulations here support a reverse phase transition. This is consistent with observations from annealing experiments conducted at T=1900 K in an argon atmosphere, where x-ray diffraction patterns suggested the formation of fragmented lonsdaleite-like layers as intermediates during the thermal degradation of diamond powders. [76] A small population of atoms with $sp^3$ hexagonal (lonsdaleite-like) local stacking is observed at the diamond–graphite interface. Given their limited abundance and spatial extent, these features should not be interpreted as a distinct or stable hexagonal diamond phase, but rather as transient local stacking variants arising during interfacial reconstruction. Here, their relative occurrence across morphologies and quench–release conditions is used as a qualitative indicator of transformation pathways, not as a quantitative measure of hexagonal diamond formation.

Previous studies have reported that rapid drops in both temperature and pressure collectively help inhibit nanodiamond graphitization. [77] In the present simulations, a two-stage cooling and depressurization process was modeled. In the first stage, a rapid temperature drops of $\Delta T_1/\Delta t = 30$ K/ps concomitate with an increased pressure release at a rate of $\Delta P_1/\Delta t = 0.590$ GPa/ps were applied until reaching a midpoint temperature of $T_{mid} = 2000$ K and pressure of $P_{mid} = 1.0$ GPa (represented by the dashed purple line, bottom trajectory, **Fig. 2**). This was followed by a second stage in which both cooling and pressure release continued at slower rates ($\Delta T_2/\Delta t = 17$ K/ps and $\Delta P_2/\Delta t = 0.005$ GPa/ps) until the system reached final conditions of $T_f = 300$ K and $P_f = 1 \times 10^{-4}$ GPa (or 1 atm). As shown in **Fig. 1D**, the resulting structure contains multiple graphitic layers, up to three layers yet remain associated with the particle surface. Here, two regions of interfacial $sp^3$ hexagonal carbon (orange). Collectively, the simulation results indicate that reductions in temperature, rather than in pressure, play a more pivotal role in inhibiting the graphitization of nanodiamond.

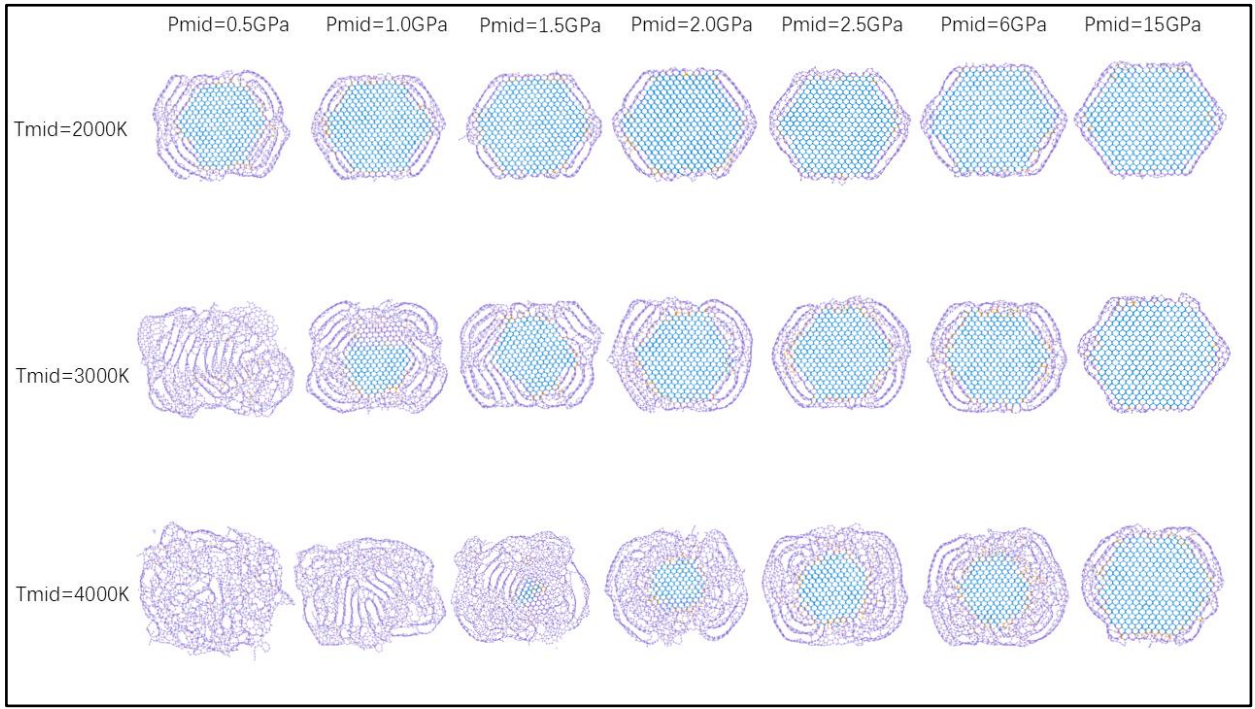

**Figure 3**. Simulated structures of the products, shown as center slices 10 Å thick, after cooling and pressure release of nanodiamonds initially adopting a cuboctahedral morphology. $P_{mid} \in \{0.5, 1.0, 1.5, 2.0, 2.5, 6, 15\}$ GPa and $T_{mid} \in \{2000, 3000,$

4000} K. Atom color is based on PTM types: cubic diamond (CUB) as blue, hexagonal diamond (HEX) as orange, and purple as $sp^2$ hybridized carbon (SP2).

The structural evolution as a function of midpoint temperature and pressure is presented in greater detail in **Fig. 3**. The final simulated structure that retains the most intact sp³-hybridized cubic diamond phase (blue atoms) corresponds to the scenario with the fastest temperature drop (row 1, $T_{mid}$ = 2000 K) and the slowest pressure-release rate ($T_{mid}$ = 2000 K; $P_{mid}$ = 15 GPa). This is consistent with the recent work of Heuser et al., who reported a combined MD simulation and experimental study on the laser-driven shock formation and release dynamics of nanodiamond synthesized from polyethylene terephthalate. [78] In brief, they demonstrated that, as temperature and pressure decrease, the diamond lattice becomes unstable, with defects initiating at the surface and propagating inward. Moreover, an accelerated temperature drop was found to suppress nanodiamond disintegration that occurs through graphitization and bond breaking. The maintenance of high-pressure during expansion is believed to enhance cooling during expansion, serving to mitigate structural disorder, and possibly promoting recrystallization upon cooling. Here, too we find the slowest pressure-release rate (right most column, **Fig. 3**) acts to preserve the $sp^3$ cubic diamond as revealed in our simulations which found the least number of graphitic layers surrounding (purple atoms) the diamond particle core (blue atoms). As the temperature cooling rate decreases (rows 2 and 3, **Fig. 3**), the slower pressure-release rates are insufficient to prevent multilayer graphitization, leading to significant disordering and delamination of the layers from the core particle

Furthermore, as shown in **Fig. 3**, the MD simulations reveal regions of metastable lonsdaleite (hexagonal diamond; orange atoms in **Fig. 3**) that emerge during the graphitization of cubic diamond. Simulations conducted under faster cooling conditions ($T_{mid}$ = 2000 K) and slow pressure release exhibit the fewest regions of hexagonal $sp^3$ carbon (lonsdaleite). In contrast, simulations performed at slower cooling rates ($T_{mid}$ = 3000 K and 4000 K) and under faster pressure-release conditions ($P_{mid}$ = 2.0–6.0 GPa) display more extensive lonsdaleite regions. Experimental observations of transitional lonsdaleite layers during thermal annealing and graphitization of cubic diamond have recently been corroborated via TEM and diffraction analysis. [76]

**Carbon nano-onions (CNOs)**. As previously discussed, nanodiamond graphitization is well established and can result in the formation of several nanocarbon phases of interest, including the so-called "bucky diamond", particles featuring a diamond core with a fullerenic shell as well as onion-like carbons (OLCs) that are also known as carbon nano-onions (CNOs), which are composed of concentric graphitic shells. [27, 79, 80] These nanophase carbons are reliably produced through the structural transformation of nanodiamond via graphitization of the diamond {111} planes at temperatures between 1300 and 1900 K. Among the intermediate structures, bucky diamond is more thermodynamically stable than CNOs. In this study, we select a nanodiamond particle with an initial octahedral crystal structure (**Fig. 1E**), as its surface is dominated by diamond {111} planes, which are known to graphitize more readily than other crystallographic planes. This allows us to investigate how varying release conditions remodel the nanodiamond into different graphitized forms. Quarter-cut views of the nanodiamond, initially possessing an octahedral crystal structure, are shown in **Fig. 4A–G** as it evolves during a simulation involving a two-stage release process. The first stage applies a rapid temperature drop ($\Delta T_1/\Delta t$ = 30 K/ps) combined with a slow pressure release ($\Delta P_1/\Delta t$ = 0.595 GPa/ps), continuing from the peak temperature and pressure until midpoint conditions are reached ($T_{mid}$ = 2000 K, $P_{mid}$ = 0.5 GPa). In the second stage, both cooling and pressure release proceed at slower rates ($\Delta T_2/\Delta t$ = 17 K/ps; $\Delta P_2/\Delta t$ = 0.005 GPa/ps), continuing until final conditions of $T_f$ = 300 K and $P_f$ = 1 × 10⁻⁴ GPa (1 atm) are achieved. At early times (90 ps, **Fig. 4B**) ca. 2 layers of *sp²* hybridized carbon (purple atoms) that are well ordered and bound to the nanodiamond {111} surfaces are observed. Furthermore, localized regions, but not fully developed interfacial layers, of *sp³*-bonded hexagonal carbon (i.e., lonsdaleite; orange atoms) are sporadically found between the *sp³* cubic diamond core and the well-formed *sp²* layers. Continued graphitization is observed between 100 ps and 105 ps (**Fig. 4C–F**), marked by the growth of additional *sp²*-hybridized carbon layers emanating from the *sp³* cubic diamond core. The outermost *sp²* carbon layers begin to exhibit signs of disorder and delamination. Overall, the resulting particle structure is consistent with experimentally observed bucky diamonds. Additionally, stabilization of the lonsdaleite-like interfacial region between the *sp³* diamond core and the surrounding *sp²* graphitic layers is also noted. Additional insights into the transition from diamond, characterized by tetrahedrally coordinated *sp³*-bonded carbon, to regions dominated by *sp²*-hybridized carbon arranged in onion-like shells can be obtained by analyzing the evolution of carbon ring statistics (**Fig. 4, panel I**). The plot reveals a sharp increase in the number of six-membered rings containing *sp²* hybridized carbon beginning around 75 ps, followed by a plateau near 100 ps. The discontinuity is ascribed to the deliberate change in the temperature and pressure-release rates in the simulations. To a lesser extent, populations of five-, seven-, and eight-membered rings are also observed; notably, not highly strained four-membered rings are formed. At the end of the simulation (200 ps) the final structure has transformed into features characteristic of a nano-onion with complete conversion to *sp²* hybridized carbon (**Fig. 4G**). Further noted is a considerable amount of disorder in the *sp²* hybridized carbon layers particularly in the core region. Through the combination of ring strain and structural disorder within the concentric carbon layers creates inherent instability of the core, leading to the transformation towards a hollow structure. [81] For example, the generation of hollow carbon nano onions through high temperature (1100 °C) annealing of nano onions produced by catalytic CVD. [82] The hollow nano onions structures have also been recovered from detonation soot. [5] Thus, the formation of a range of quasi-spherical particles ranging from bucky-diamond to carbon nano onions to hollow core-shell particles can all be produced from the graphitization of nanodiamond.

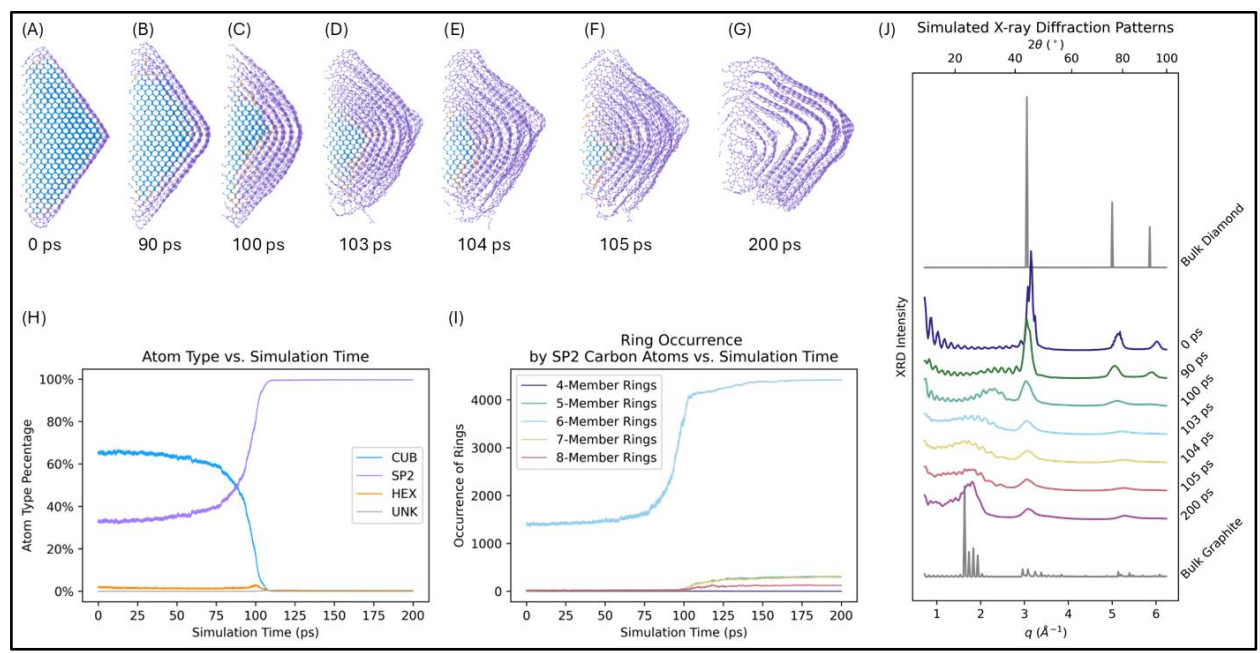

**Figure. 4** (A) – (G) Quarter-cut views of the nanodiamond with an octahedral crystal structure, showing the evolution of graphitization at selected time points during the simulation. T-P conditions used in simulation: $\Delta T_1/\Delta t$ = 30 K/ps, $\Delta P_1/\Delta t$ = 0.595 GPa/ps, $\Delta T_2/\Delta t$ = 17 K/ps, $\Delta P_2/\Delta t$ = 0.005 GPa/ps. Atom color is based on PTM types: cubic diamond (CUB) as blue, hexagonal diamond (HEX) as orange, and $sp^2$ hybridized state (SP2) as purple. (H) Number of carbon atoms categorized by local structure type. 65.5% CUB type atoms start at the beginning of the simulation, and 0.4 % at the end. (I) Frequency of N-member rings formed by $sp^2$-hybridized carbon atoms. $sp^2$ type atoms form 1,403 6-member rings at the beginning of the simulation, and 4,419 at the end. (J) Simulated X-ray diffraction patterns at selected time points during the simulation. X-ray diffraction pattern (Cu Ka) shows a trend of diamond structure transitioning to graphite structure.

**Carbon Dots (Cdots):** Recently, there has been growing interest in carbon dots (Cdots), which are zero-dimensional, carbon-based nanomaterials typically less than 10 nm in diameter. Of particular interest are fluorescent Cdots, which unlike traditional semiconductor quantum dots, exhibit non-blinking, stable fluorescence, are biocompatible, have low toxicity, and are more environmentally benign. These advantages make them attractive alternatives to toxic II–VI (Cd-based) and IV–VI (Pb-based) chalcogenide quantum dots. [82, 83] Cdots are often composed of fragments of graphene sheets or graphite, and in some cases are referred to as graphene quantum dots (GQDs) due to their structural similarity. Fluorescent, quasi-spherical graphene particles (Cdots) can be synthesized through chemical oxidation of hollow carbon nano-onions (CNOs). [84] These hollow CNOs are considered ideal precursors due to their intrinsic defects and porosity, which render them susceptible to acid-induced oxidation and cage-opening of their fullerenic carbon layers, composed of mixed hexagonal and pentagonal rings. [86] This process leads to the exfoliation of graphene sheets, which may subsequently form GQDs. Under conditions such as high-explosive detonations, an open question remains: Are Cdots formed from graphitized nano-onions, from hollow CNOs, or directly from detonation nanodiamond? Molecular dynamics simulations tracking the temporal evolution of six-membered carbon rings during the transformation from nanodiamond to nano-onions indicate that, in the initial graphitization stage, the structure evolves into carbon sheets rich in hexagonal rings (see **Fig. 4, panel I**). At the second temperature and pressure drop stage, there is an observed increase in five-, seven-, and eight-membered rings, suggesting that the outer graphitized layers are dominated by hexagonal carbon, while the more curved inner regions require incorporation of non-hexagonal rings (e.g., pentagons) to maintain curvature. As the carbon shells grow inward, increasing curvature and strain may eventually lead to the destabilization and loss of inner layers, resulting in the formation of hollow core–shell structures. Hollow core–shell nanocarbons have previously been reported by Huber et al. from detonations of Composition B. [5] It is possible that the loss of the highly structurally defective interior layers causes the sheets to undergo further cleavage and exfoliation into graphene sheets, which then self-organize into graphene dots.

Alternatively, we propose a different pathway: the direct growth of Cdots from nanodiamonds, adopting a morphology that promotes the formation of the graphite-like layers characteristic of Cdots. The nanodiamond crystal morphology that consistently graphitized into layered structures characteristic of Cdots was the hexagonal prism (**Fig. 4, panel I**). A hexagonal prismatic nanodiamond structure was generated by truncating the {111} vertices of a cube, followed by truncation of the {110} edges. Simulations were carried out using a two-stage temperature-drop and pressure-release procedure. In the first stage, a rapid temperature drop ($\Delta T_1/\Delta t$ = 30 K/ps) was combined with a slow pressure release ($\Delta P_1/\Delta t$ = 0.595 GPa/ps) from the peak temperature and pressure, continuing until midpoint conditions were reached ($T_{mid}$ = 2000 K, $P_{mid}$ = 0.5 GPa). In the second stage, both cooling and pressure release proceeded at slower rates ($\Delta T_2/\Delta t$ = 17 K/ps; $\Delta P_2/\Delta t$ = 0.005 GPa/ps), continuing until the final conditions were reached ($T_f$ = 300 K, $P_f$ = 1 × 10$^{-4}$ GPa, or 1 atm). At early simulation times (0–90 ps, **Fig. 5A, B**), up to three layers of graphite formed on the {111} planes. Some graphitization was also observed on the top and bottom {110} planes, although this did not result in layered structures as seen on the {111} planes. As the graphitization process progressed (100–120 ps, **Fig. 5C–E**), graphite layers continued to grow, maintaining a parallel orientation toward the particle interior, which remained predominantly cubic diamond. At approximately 110 ps (**Fig. 5E**), a time point following the transition to the second-stage, slower cooling and pressure-release conditions, the cubic diamond core disintegrated, with no remaining *sp³* cubic (blue) diamond atoms in the center and only polarized fragments localized near the {110} surfaces of the particle. Notably, the beginning of the second-stage cooling and pressure release marked the point of maximum *sp³* hexagonal carbon content (orange). The final structure, captured at 200 ps (**Fig. 5G**), contained only sporadic *sp³* carbon atoms. Most of the particle had fully converted to sp² carbon, and its morphology had evolved into a highly ordered, quasi-spherical structure characterized by regions of layered sp² carbon. The evolution of both carbon hybridization and ring member content is summarized in **Fig. 5H and I**, respectively. Although beyond the scope of the current work, we conjecture that a hexagonal prismatic nanodiamond particle could be formed through an anisotropic shock wave. Previous studies on laser machining of crystalline diamond have shown that different crystallographic planes exhibit varying ablation thresholds, with the {100} plane demonstrating the highest resistance to ablation. [81] Thus, it would be interesting to investigate whether anisotropic shock waves could be used to produce novel, non-native crystal morphologies of nanodiamond.

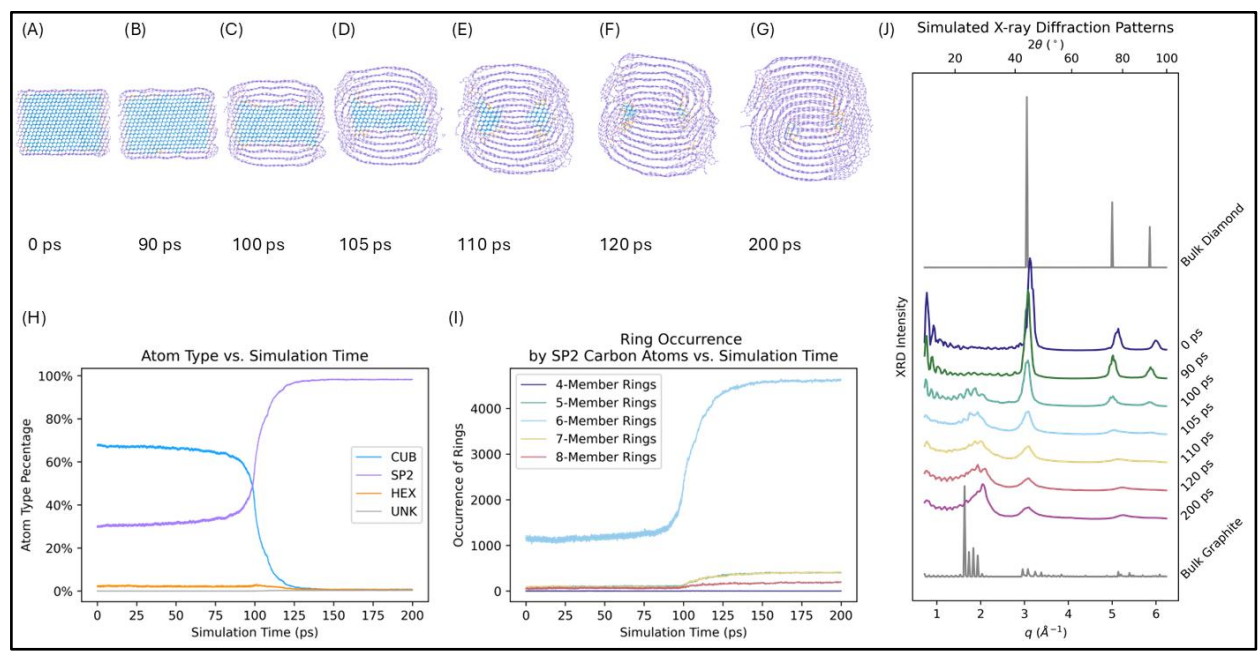

**Figure 5.** (A)- (G) Side views of the hexagonal prism nanodiamond showing the evolution of graphitization at selected time points during the simulation. Atom color is based on PTM types: cubic diamond (CUB) as blue, hexagonal diamond (HEX) as orange, and $sp^2$ hybridized state (SP2) as purple. (H) Number of carbon atoms categorized by local structure type. 67.7% CUB type atoms start at the beginning of the simulation, and 0.7 % at the end. (I) Frequency of N-member rings formed by $sp^2$-hybridized carbon atoms. $sp^2$ type atoms start at 29.9% (1,129 6-member rings) at the beginning of the simulation and evolve to 98.3 % (4,621 6-member rings) at the end. (J) Simulated X-ray diffraction patterns at selected time points during the simulation. X-ray diffraction pattern (Cu Kα) shows a trend of diamond structure transitioning to graphite structure.

**T-P variations impact on graphitization product**. The work outlined above examined nonlinear quench and pressure-release schemes in which temperature decreased more rapidly than pressure. To more fully understand the interplay between temperature-quench rate and pressure-release rate, we performed a series of simulations on octahedral nanodiamond in which the intermediate temperature ($T_{mid}$) was varied to 4000 K and 3000 K, in addition to the previously discussed 2000 K. These simulations were conducted across a range of intermediate pressures ($P_{mid}$): 15 GPa, 6 GPa, 2.5 GPa, 2.0 GPa, 1.5 GPa, 1.0 GPa, and 0.5 GPa (**Fig. 6**). As shown in **Fig. 6** (lower left quadrant), a slow temperature drop ($T_{mid}$ = 4000 K) combined with a rapid pressure release ($P_{mid}$ = 0.5 GPa) results in complete amorphization of the octahedral diamond nanocrystal (see **Fig. 1E**). In this case, the entire nanodiamond particle is converted from $sp^3$ cubic diamond to highly disordered $sp^2$ carbon, with a conversion rate exceeding 99%. In contrast, a slow temperature drop ($T_{mid}$ = 4000 K) coupled with a slow pressure release ($P_{mid}$ = 15 GPa) yields a particle that retains a $sp^3$ cubic diamond core, surrounded by multilayers of well-ordered $sp^2$ carbon and a discernible interfacial layer of $sp^3$ hexagonal diamond (lonsdaleite), as shown in **Fig. 6**, lower right quadrant. At the other extreme (**Fig. 6**, upper right quadrant), fast temperature reduction ($T_{mid}$ = 2000 K) coupled with a slow pressure release ($P_{mid}$ = 15 GPa) leads to minimal particle surface graphitization with maximum retention of $sp^3$ cubic diamond. A rapid thermal quench ($T_{mid}$ = 2000 K) in conjunction with an abrupt pressure release ($P_{mid}$ = 0.5 GPa) yields a structurally ordered nano-onion morphology, exhibiting discernible interlayer spacing modulations (undulatory distortions), as illustrated in **Fig. 6** (upper left quadrant).

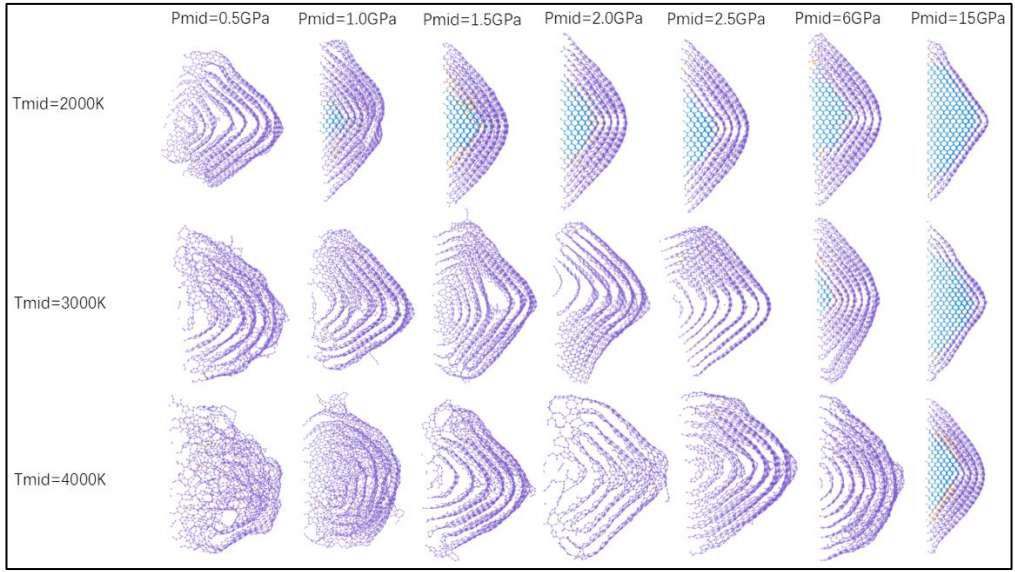

**Figure 6**. Quarter cut view of the quench products of octahedron nanodiamond at $P_{mid} \in$ {0.5, 1.0, 1.5, 2.0, 2.5, 6, 15} GPa and $T_{mid} \in$ {2000, 3000, 4000} K. Atom color is based on PTM types: cubic diamond (CUB) as blue, hexagonal diamond (HEX) as orange, and $sp^2$ hybridized as purple.

A series of similar T-P variation simulations were also performed with the hexagonal prismatic nano crystal (**Fig. 7**). Briefly, the fast temperature quench ($T_{mid}$ 2000 K) retains the most diamond core over a wide range of pressure releases ($P_{mid}$ = 1.0 GPa – 15 GPa) and at the fastest pressure release ($P_{mid}$ = 0.5 GPa) results in a particle possessing well ordered $sp^2$ carbon layers. Alternatively, slow quench rates over a wide range of pressure releases ($P_{mid}$ = 1.5 GPa – 6 GPa) rapidly graphitize into particles containing ordered layers of $sp^2$ carbon sheets. As expected, slow temperature drops coupled with fast pressure release results in highly disordered particle structure.

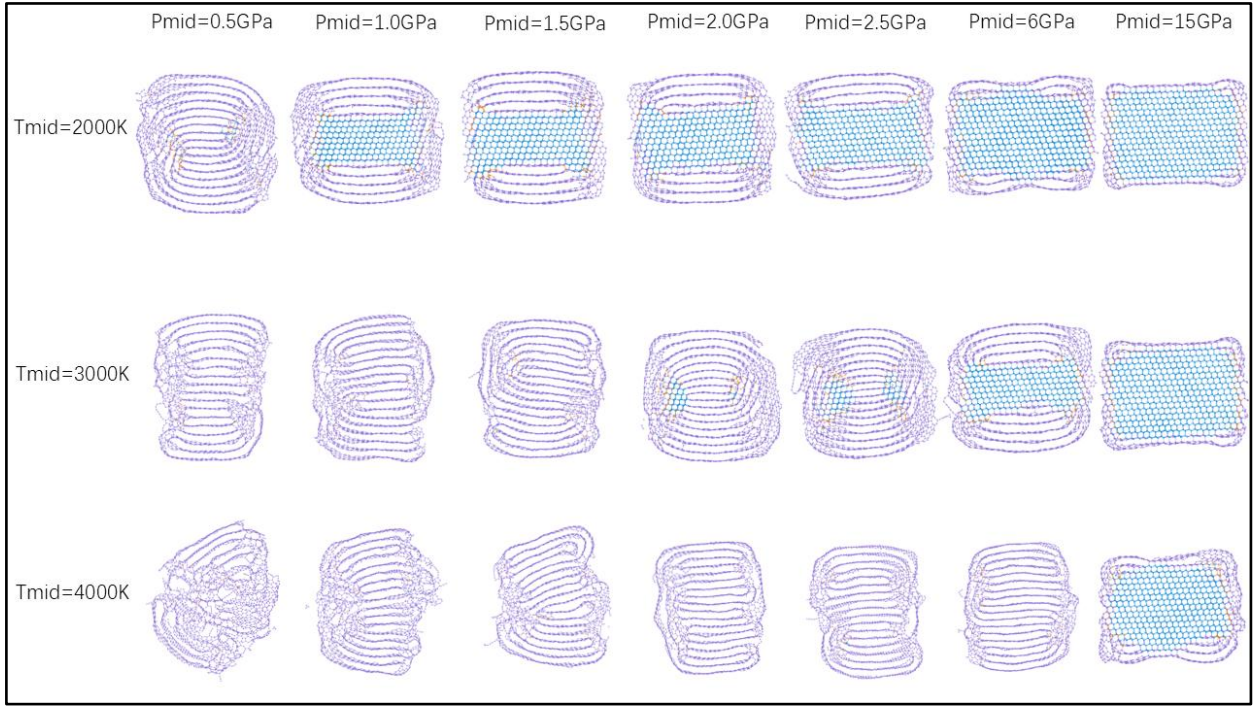

**Figure 7**. Center slice of quenched hexagonal prism nanodiamond at $P_{mid} \in \{0.5, 1.0, 1.5, 2.0, 2.5, 6, 15\}$ GPa and $T_{mid} \in \{2000, 3000, 4000\}$ K. Atoms are colored by PTM type: cubic diamond (CUB) in blue, hexagonal diamond (HEX) in orange, and $sp^2$-hybridized in purple. The prism is oriented with {111} side faces and {110} top and bottom faces.

MD simulations reveal that nanodiamond facet energetics play a decisive role in determining the resulting nanophase carbon structures. Octahedral nanodiamonds, dominated by low-energy {111} facets, preferentially undergo surface-driven graphitization through the conversion of diamond {111} planes into graphite {002} planes. This pathway is favored by both energetic and kinetic considerations: the {111} surface has the lowest reconstruction energy, as its six-membered rings can flatten into graphitic sheets with minimal bond breaking, while interface migration proceeds via a cooperative, zipper-like mechanism in which three adjacent {111} planes transform into two graphitic layers, lowering the activation barrier and preserving lattice continuity [87]. As graphitization progresses, curved $sp^2$ reconstructions form via concentric shell growth, efficiently accommodating curvature and topological defects within discrete layers and minimizing strain. At the atomistic level, the emerging diamond–graphite interface exhibits topotactic relationships between diamond {111} facets and the nascent graphite layers, often accompanied by stacking disorder and transient lonsdaleite-like motifs [88]. These interfacial features are consistent with high-resolution experimental observations of diamond–graphite and diamond–graphene composites, which report atomically sharp yet structurally complex $sp^3$–$sp^2$ junctions containing stacking faults, hexagonal diamond intermediates, and coherently aligned graphitic layers. Similar mixed-bonding interfaces, curvature-driven graphitic shells, and partially ordered $sp^2$ layers adjacent to diamond cores have also been observed in thermally or shock-treated diamond systems [83–85]. Ultimately, the extent of graphitization, governed by the quench–release conditions, dictates whether the resulting particles adopt bucky-diamond, onion-like, or hollow core–shell morphologies.

In contrast, hexagonal prismatic nanodiamonds expose extended {100} and {110}-like facets and edges, which direct anisotropic stress relaxation and promote directional $sp^2$ nucleation. These facets favor planar or weakly curved $sp^2$ domains, yielding stacked, layered graphitic motifs characteristic of carbon dots (C-dots) and related nanocarbon systems, rather than concentric closed shells. Collectively, these observations support a nonequilibrium transformation framework in which morphology-encoded anisotropic stress and quench–release pathways, rather than equilibrium phase boundaries alone, govern carbon transformation outcomes.

**Machine Learning Model for Graphitization Layer Prediction**. Four different ML models; B-spline, random forest (RF), gradient boosting, and multilayer perceptron (MLP) were evaluated by distilling data from over 10,000 node-hours of ReaxFF simulations performed on an exascale supercomputer, to determine which model best captures the nonlinear thermodynamic pathways governing nanodiamond graphitization. These models focus on octahedral and hexagonal prism geometries, which are of particular interest because they give rise to nano-carbon onions and carbon dots during structural evolution. Within these predictive models, the maximum number of graphite layers is capped at ten, a direct consequence of the finite dimensions of the nanodiamonds under study. Accuracy metrics, including $R^2$ and mean squared error (MSE), are summarized in **Table 3**. The MLP model achieves good predictive accuracy, with $R^2 = 0.893$ on the training set and $R^2 = 0.904$ on the test set, as summarized in **Table 3** and further illustrated in **Figure 8A**. We note that traditional nonlinear regression approaches were also explored as baselines for mapping the 4D input tensor to graphitization metrics. These models exhibited limited predictive capability, particularly in capturing the non-monotonic and regime-dependent responses associated with different quench and pressure-release pathways and performed poorly on held-out trajectories. In contrast, the ML models reported here achieve consistent predictive accuracy across unseen simulations and provide interpretable sensitivity to thermodynamic history, motivating their use over simpler nonlinear fitting approaches. We have released all the scientific software needed to reproduce these results, along interactive visualizations in the GitHub repository: https://github.com/williamyxl/nanodiamond.

**Table 3**. Summary of performance metrics for machine learning models evaluated in this study

| Performance metrics | B-spline | Gradient Boosting | Random Forest | Multilayer Perception |
|---|---|---|---|---|
| $R^2$ training | 0.823 | 0.809 | 0.970 | 0.893 |
| MSE training | 2.19 | 2.38 | 0.368 | 1.33 |
| $R^2$ testing | 0.838 | 0.881 | 0.946 | 0.904 |
| MSE testing | 1.94 | 2.15 | 0.653 | 1.15 |

A 3D surface plot of the regression model provides a direct visualization of the MLP model's predictions based on the two input variables, T and P, and the corresponding number of graphitic layers formed (**Fig. 8B**). Across the explored range of 1000–5000 K and 1 atm–60 GPa, the model extrapolates predictions for structures containing between zero and ten graphitic layers. The MLP model captures a smooth, continuous surface across the P–T space. The surface plot highlights regions of minimal or no graphitization in the lower-left quadrant, corresponding to low intermediate temperature ($T_{mid}$) and high intermediate pressure ($P_{mid}$). In contrast, maximum graphitization is observed in the upper-right quadrant, where $T_{mid}$ is high and $P_{mid}$ is low. Of particular interest is the region where the surface slope becomes minimal (purple region), found when $P_{mid}$ exceeds 2 GPa. This suggests that graphitization is unlikely at $P_{mid} > 2$ GPa. Alternatively full graphitization is found at the surface plot flat plateau at conditions of $P_{mid} < 2$ GPa and $T_{mid} > 3000$ K (yellow region). The steeper transition region on the surface contour plot (green region) reveals a zone of very high sensitivity to the coupled effects of T and P and coincides with the absence of simulation data for the 5–8 graphitic layers. The absence of these layers in the predicted phase space warrants further investigation using refined decompression rates and finer simulation time steps and will be the focus of future studies. Because the initial and final thermodynamic states ($P_0, T_0$) and ($P_f, T_f$) are fixed in the present dataset, the four components of the input tensor are partially correlated; nevertheless, retaining all four parameters provides a compact description of the full quench–release trajectory, while Fig. 8 demonstrates that ($P_{mid}, T_{mid}$) capture the dominant contribution to the observed graphitization trends.

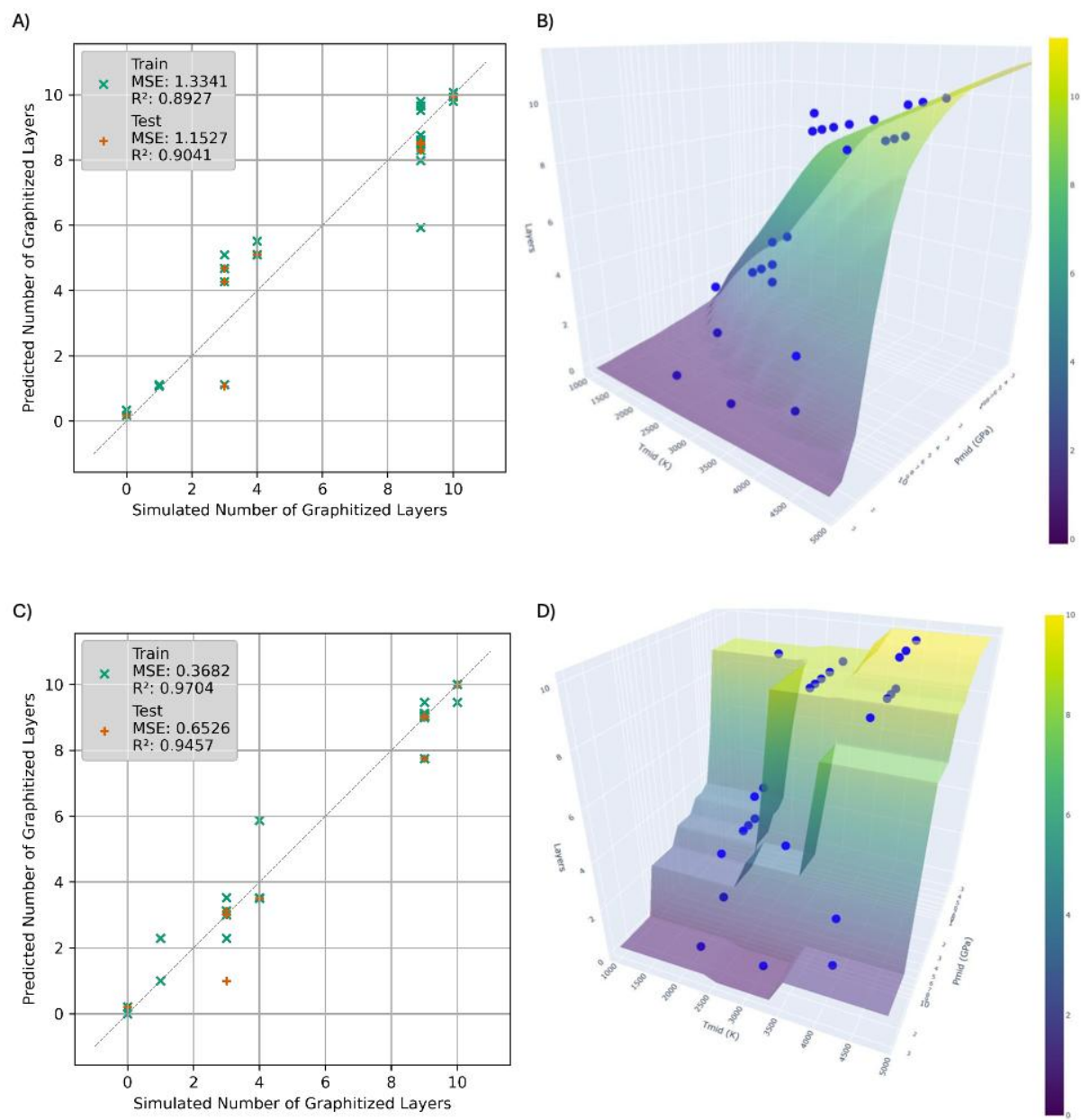

**Figure 8.** (A) Comparison between the number of graphitic layers produced in molecular dynamics simulations and the predictions from the MLP model. Training data are shown as green "×" markers, and testing data are shown as orange "+" markers. The solid grey line serves as a visual reference for perfect agreement. (B) 3D surface plot generated using the MLP model trained on MD simulation data, illustrating predicted graphitization as a function of temperature and pressure. The color bar indicates the number of graphitized layers, purple is low, yellow is high. $T_{mid}$ (K) is in linear scale, and $P_{mid}$ (GPa) is in log scale. Blue scatter points are the data acquired from ReaxFF simulations. (C) Comparison between the number of graphitic layers produced in molecular dynamics simulations and the predictions from the random forest model. Training data are shown as green "×" markers, and testing data are shown as orange "+" markers. The solid grey line serves as a visual reference for perfect agreement. (D) 3D surface plot generated using the random forest model trained on MD simulation data, illustrating predicted graphitization as a function of temperature and pressure. The color bar indicates the number of graphitized layers, purple is low, yellow is high. $T_{mid}$ (K) is in linear scale, and $P_{mid}$ (GPa) is in log scale. Blue scatter points are the data acquired from ReaxFF simulations.

The RF model demonstrates higher predictive accuracy, achieving $R^2 = 0.970$ on the training set and $R^2 = 0.946$ on the test set than the MLP model (**Fig. 8C**). The relationship between the predicted number of graphitic layers and the midpoint values in P–T space yields a 3D contour surface that is rough and stepped, indicating potential overfitting and yielding non-physical interpretations of the relationship between the process parameters ($P_{mid}$, $T_{mid}$) and the resulting graphitic layer formation (**Fig. 8D**). This model employs 100 decision trees (estimators), each with a maximum depth of 8. As a result, the number of fitting parameters is significantly greater than that of the MLP model, indicating a more complex and potentially less generalizable model. Thus, the MLP model provides a rapid and predictive capability for identifying the conditions that lead to varying degrees of nanodiamond graphitization, offering a significant speed advantage over the time-consuming and computationally intensive exascale MD simulations. Instead of waiting two weeks for a single trajectory to complete, the graphitization outcome for a given temperature–pressure path can now be estimated in a fraction of a second. This enables computationally guided design of exotic carbon nanostructures.

Overall, this work represents the first application of ML to enable rational selection of extreme thermodynamic conditions for tailoring nanodiamonds into novel and functional carbon nanostructures. Future work will focus on incorporating the initial nanodiamond crystal structure (e.g., morphology) into the model, with the goal of more fully predicting how specific conditions can be used to produce desired nanophase carbon materials.

**Conclusions**

Herein we present a comprehensive atomistic modeling framework to elucidate the influence of nanodiamond morphology, specifically cuboctahedral, octahedral, and hexagonal prismatic geometries, combined with nonlinear thermal quenching and pressure release trajectories originating from peak conditions ($T_0$ = 5000 K, $P_0$ = 60 GPa), entirely within the diamond stability region of the carbon phase diagram, on the formation pathways of novel nanostructured carbon phases. Diamond graphitization studies both experimentally and theoretically have mostly centered on heat treatment of diamond. [89] In contrast to previous studies, this model explicitly incorporates pressure as an independent variable, enabling detailed investigation of the coupled effects of temperature and pressure on structural evolution. Collectively, across all starting nanodiamond particle morphologies, rapid temperature quenching combined with gradual pressure release tends to preserve $sp^3$-hybridized carbon, stabilizing cubic diamond and suppressing or delaying surface graphitization. In contrast, slow temperature reduction coupled with rapid pressure release accelerates surface graphitization, particularly on the {111} facets, resulting in considerable disorder within the $sp^2$-hybridized carbon layers.

The results presented here establish a conservative baseline for understanding graphitization pathways under dynamic quench–release conditions and motivate several important directions for future work. Previous studies under static compression have demonstrated that diamond graphitization depends strongly on the applied stress state, with non-hydrostatic loading enabling transformation at substantially lower temperatures and pressures than quasi-hydrostatic conditions. [90] By comparison, experimental investigations of diamond graphitization under dynamic compression remain limited. In detonation environments, loading is inherently non-hydrostatic and accompanied by large shear and deviatoric stresses, suggesting that mechanochemical effects may further lower the effective barriers for $sp^3$–$sp^2$ transformations. [91] By deliberately isolating hydrostatic quench–release effects, the present simulations disentangle thermodynamic driving forces from kinetic pathway selection; as a result, the transformation thresholds reported here should be interpreted as conservative. Incorporating explicit shear loading is therefore expected to further bias transformation pathways and represents a critical avenue for future investigation.

In addition, the present simulations neglect bulk and surface defects as well as variations in surface chemistry, all of which are known to influence carbon phase stability. Structural defects such as vacancies, grain boundaries, and topological irregularities act as local stress concentrators and nucleation sites for bond rearrangement, thereby reducing the pressures and temperatures required for $sp^3 \leftrightarrow sp^2$ transformation. [92, 93] Surface termination effects are likewise expected to shift the absolute onset of graphitization and diamond stabilization relative to idealized bare-surface models. Reactive MD studies of hydrogenated multilayer graphene demonstrate large reductions in transformation stress and strain, along with the emergence of metastable or stable diamond films following unloading. Consequently, surface passivation may bias transformation thresholds toward enhanced $sp^3$ retention compared with bare-surface simulations. Accordingly, these results define relative mechanistic trends across quench and pressure-release pathways. [94]

This work does not exhaustively explore nanodiamond particle size beyond the limited range of 2–4 nm, which emphasizes surface and curvature effects relative to larger particles. Cooling and pressure-release rates occur on picosecond timescales, as dictated by MD simulations, and therefore do not reproduce absolute experimental kinetics. In addition, the use of a classical ReaxFF description imposes known limitations on quantitative energetics, and the simulations intentionally neglect chemical impurities and surface functionalization commonly present in detonation environments. These factors may shift absolute transformation thresholds; however, they do not alter the central conclusions of this work, which focus on relative mechanistic trends and morphology-dependent pathway selection under controlled thermodynamic histories. Collectively, systematic variation of stress state, defect content, surface chemistry, and particle size defines a rich space for future investigation, particularly for understanding how these parameters may be leveraged to tune nanocarbon products that can be reproducibly generated under controlled graphitization conditions relevant to detonation nanodiamond.

This work further explores how the initial nanodiamond morphology, under thermal quenching and slow pressure release conditions conducive to graphitization, can be remodeled into novel nanostructured particles of technological interest. Specifically, bucky diamonds are generated from either cuboctahedral or octahedral starting nanodiamonds; nano-onions result from extended graphitization of octahedral particles; and with continued structural evolution (graphitization), hollow core–shell graphitic particles can form. While it is hypothesized that further disintegration of these hollow core–shell graphitic structures may yield graphene / graphite fragments that subsequently reassemble into stacks of *sp²*-hybridized carbon, ultimately forming carbon dots (Cdots), we also simulate the controlled formation of Cdots via temperature–pressure drops applied to nanodiamonds with a hexagonal prismatic morphology. Mechanistically, this transformation pathway may also explain the formation of diamond–graphite nanocomposites, such as diaphite, which have been recovered from detonation soot or found naturally in meteorites. [95-97] These hybrid or nanocomposite structures exhibit interspersed domains of *sp³*-bonded diamond- and *sp²*-bonded graphene layers like the composite materials generated by simulation during the nonlinear temperature and pressure drops for the hexagonal prism. Lastly, the simulations also delineate T-P conditions that favor the stabilization of hexagonal diamond (lonsdalite). Specifically, simultaneous slow temperature and pressure reductions were found to most favor the formation of hexagonal diamond. While little work has focused on studying the conversion of cubic diamond into graphite, studies examining the conversion of graphite to hexagonal diamond under shock compression has been reported. [75]. It is believed that under shock compression the graphite basal plane slides and puckers to form diamond. [98]

A significant advancement of this work lies in translating extensive reactive MD data, exceeding $10^5$ node-hours of computation into a predictive ML framework for nanocarbon synthesis design. By training supervised regression models on curated trajectory data, we establish quantitative mappings between nonlinear quench–release profiles and graphitization outcomes, expressed as the number of *sp²* layers formed. Among the evaluated regressors (MLP, B-spline, RF, and gradient boosting) the MLP and RF models achieved the highest predictive accuracy, with $R^2$ values exceeding 0.90 and 0.94, respectively, and mean squared errors below 1.2. Moreover, the MLP model captures a smooth, nonlinear dependence of graphitization on both intermediate temperature ($T_{mid}$) and pressure ($P_{mid}$), identifying low-pressure (<2 GPa), moderate-temperature (2000–3000 K) regimes as optimal for controlled $sp^2$-layer growth. In contrast, while the RF model achieves higher overall accuracy, it produces steeper and less physically continuous prediction surfaces, indicative of overfitting to localized regions in parameter space. These results underscore the robustness of neural network–based regression for generalizing across untested thermodynamic pathways and highlight its potential for the inverse design of carbon nanostructures.

In summary, by compressing massive atomistic datasets into compact, predictive models, this ML framework transforms exascale simulation outputs into near-instantaneous property estimators. What once required weeks of high-performance computation can now be evaluated in seconds, enabling rapid screening of synthesis conditions and guiding experimental efforts toward desired nanocarbon morphologies. To our knowledge, this represents the first quantitative prediction capability for estimating the number of graphitic layers formed in nanodiamond-derived materials. Moreover, the framework is inherently generalizable: incorporating additional morphological descriptors (e.g., surface curvature, facet composition) and dynamic observables (e.g., strain rate, local disorder metrics) can further enhance model fidelity and extend applicability to a wider range of carbon and hybrid nanomaterials. The broader implications of this work extend beyond post-detonation carbon synthesis, offering insights into phase stability and carbon transformations under extreme conditions relevant to planetary science, astrochemistry, and high-pressure materials processing. Future efforts will focus on scaling the framework to chemically heterogeneous systems and incorporating coupled shock–release dynamics to more accurately represent detonation environments. By unifying mechanistic insight, morphology-specific synthesis pathways, and predictive modeling, this approach provides a powerful foundation for the rational design of detonation-driven carbon nanomaterials for high-value applications in energy storage, quantum sensing, and biomedicine.

**Acknowledgements**

MAF, SC, EH, MK designed research; XY conducted research; MAF, XY analyzed data; and MAF, XY, EH wrote the paper. EH had primary responsibility for final content. All authors read and approved the final manuscript. The authors acknowledge fruitful discussions with Álvaro Vázquez-Mayagoitia.

Funding: This work was supported by the Laboratory Directed Research and Development (LDRD) for Argonne National Laboratory, provided by the Director, Office of Science, of the U.S. Department of Energy under contract no. DE-AC02-06CH11357. This research used resources of the Argonne Leadership Computing Facility, which is a U.S. Department of Energy Office of Science User Facility operated under contract DE-AC02- 06CH11357. EH acknowledges support from NSF grants OAC-2514142 and OAC-2209892. An award for computer time was provided by the U.S. Department of Energy's Innovative and Novel Computational Impact on Theory and Experiment (INCITE) Program. This research used supporting resources at the Argonne and the Oak Ridge Leadership Computing Facilities. The Oak Ridge Leadership Computing Facility at the Oak Ridge National Laboratory is supported by the Office of Science of the U.S. DOE under Contract No. DE-AC05-00OR22725. This research used both the DeltaAI advanced computing and data resource, which is supported by the National Science Foundation (award OAC 2320345) and the State of Illinois, and the Delta advanced computing and data resource which is supported by the National Science Foundation (award OAC 2005572) and the State of Illinois. Delta and DeltaAI are joint efforts of the University of Illinois Urbana-Champaign and its National Center for Supercomputing Applications.

**References**

1. Nasir, S., Hussein, M. Z., Zainal, Z. & Yusof, N. A. Carbon-Based Nanomaterials/Allotropes: A Glimpse of Their Synthesis, Properties and Some Applications. *Materials (Basel)* **11**, 295-319 (2018). https://www.mdpi.com/1996-1944/11/2/295
2. Georgakilas, V., Perman, J. A., Tucek, J. & Zboril, R. Broad Family of Carbon Nanoallotropes: Classification, Chemistry, and Applications of Fullerenes, Carbon Dots, Nanotubes, Graphene, Nanodiamonds, and Combined Superstructures. *Chem. Rev.* **115**, 4744-4822 (2015). https://doi.org/10.1021/cr500304f
3. Greiner, N. R., Phillips, D. S., Johnson, J. D. & Volk, F. Diamonds in detonation soot. *Nature* **333**, 440-442 (1988). https://doi.org/10.1038/333440a0
4. Bagge-Hansen, M. et al. Detonation synthesis of carbon nano-onions via liquid carbon condensation. Nat Commun 10, 3819-3827 (2019). https://doi.org/10.1038/s41467-019-11666-z
5. Huber, R. C. *et al.* Extreme condition nanocarbon formation under air and argon atmospheres during detonation of composition B-3. *Carbon* **126**, 289-298 (2018). https://doi.org/10.1016/j.carbon.2017.10.008
6. Hui, Y. Y., Cheng, C.-A., Chen, O. Y. & Chang, H.-C. in *Carbon Nanoparticles and Nanostructures* (eds Nianjun Yang, Xin Jiang, & Dai-Wen Pang) 109-137 (Springer International Publishing, 2016).
7. Zeiger, M., Jäckel, N., Mochalin, V. N. & Presser, V. Review: carbon onions for electrochemical energy storage. *J. Mater. Chem. A.* **4**, 3172-3196 (2016). https://doi.org/10.1039/C5TA08295A
8. El-Toni, A. M. *et al.* Design, synthesis and applications of core–shell, hollow core, and nanorattle multifunctional nanostructures. *Nanoscale* **8**, 2510-2531 (2016). https://doi.org/10.1039/C5NR07004J
9. Amin, M., Mottalebizadeh, A. & Borji, S. Influence of cooling medium on detonation synthesis of ultradispersed diamond. *DRM* **18**, 611–614 (2009). https://doi.org/10.1016/j.diamond.2008.10.048
10. Christenson, J. G. *et al.* The role of detonation condensates on the performance of 1,3,5-triamino-2,4,6-trinitrobenzene (TATB) detonation. *J. Appl. Phys* **132** (2022). https://doi.org/10.1063/5.0091799
11. Hammons, J. A. *et al.* Observation of Variations in Condensed Carbon Morphology Dependent on Composition B Detonation Conditions. *Propellants Explos. Pyrotech.* **45**, 347-355 (2020). https://doi.org/10.1002/prep.201900213
12. Mykhaylyk, O. O., Solonin, Y. M., Batchelder, D. N. & Brydson, R. Transformation of nanodiamond into carbon onions: A comparative study by high-resolution transmission electron microscopy, electron energy-loss spectroscopy, x-ray diffraction, small-angle x-ray scattering, and ultraviolet Raman spectroscopy. *J. Appl. Phys.* **97**, 074302 (2005). https://doi.org/10.1063/1.1868054
13. Ducrozet, F. *et al.* New Insights into the Reactivity of Detonation Nanodiamonds during the First Stages of Graphitization. *Nanomaterials* **11 (10)**, 2671-2687 (2021). https://doi.org/10.3390/nano11102671
14. Firestone, M. A. *et al.* Structural Evolution of Detonation Carbon in Composition B by X-ray Scattering. *Aip Conf Proc* **1793,** 030010 (2017). https://doi.org/Artn
15. van Dishoeck, E. F. Astrochemistry: overview and challenges. *Proc. Int. Astro. Union* **13**, 3-22 (2017). https://doi.org/10.1017/S1743921317011528
16. Handley, C. A., Lambourn, B. D., Whitworth, N. J., James, H. R. & Belfield, W. J. Understanding the shock and detonation response of high explosives at the continuum and meso scales. *App. Phys. Rev.* **5**, 011303 (2018). https://doi.org/10.1063/1.5005997
17. Oganov, A., Hemley, R., Hazen, R. & Jones, A. Structure, Bonding, and Mineralogy of Carbon at Extreme Conditions. *Rev. Mineral. Geochem.* **75**, 47-77 (2013). https://doi.org/10.2138/rmg.2013.75.3

18  Apostolova, T., Kurylo, V. & Gnilitskyi, I. Ultrafast Laser Processing of Diamond Materials: A Review. *Front. Phys.* **9**, 650280 (2021). https://doi.org/10.3389/fphy.2021.650280
19  Frauenheim, T., Blaudeck, P., Stephan, U. & Jungnickel, G. Atomic structure and physical properties of amorphous carbon and its hydrogenated analogs. *Phys. Rev. B* **48**, 4823-4834 (1993). https://doi.org/10.1103/PhysRevB.48.4823
20  Porezag, D., Frauenheim, T., Köhler, T., Seifert, G. & Kaschner, R. Construction of tight-binding-like potentials on the basis of density-functional theory: Application to carbon. *Phys. Rev. B* **51**, 12947-12957 (1995). https://doi.org/10.1103/PhysRevB.51.12947
21  Barnard, A. S., Russo, S. P. & Snook, I. K. Structural relaxation and relative stability of nanodiamond morphologies. *DRM* **12**, 1867-1872 (2003). https://doi.org/10.1016/S0925-9635(03)00275-9
22  Barnard, A. S., Russo, S. P. & Snook, I. K. Coexistence of bucky diamond with nanodiamond and fullerene carbon phases. *Phys. Rev. B* **68**, 073406 (2003). https://doi.org/10.1103/PhysRevB.68.073406
23  Barnard, A. S. & Sternberg, M. Crystallinity and surface electrostatics of diamond nanocrystals. *J. Mater. Chem.* **17**, 4811-4819 (2007). https://doi.org/10.1039/B710189A
24  Bródka, A., Hawełek, Ł., Burian, A., Tomita, S. & Honkimäki, V. Molecular dynamics study of structure and graphitization process of nanodiamonds. *J. Mol. Struct.* **887**, 34-40 (2008). https://doi.org/10.1016/j.molstruc.2008.01.055
25  Bródka, A., Zerda, T. W. & Burian, A. Graphitization of small diamond cluster — Molecular dynamics simulation. *DRM* **15**, 1818-1821 (2006). https://doi.org/10.1016/j.diamond.2006.06.002
26  Leyssale, J. M. & Vignoles, G. L. Molecular dynamics evidences of the full graphitization of a nanodiamond annealed at 1500K. *Chem. Phys. Lett.* **454**, 299-304 (2008). https://doi.org/10.1016/j.cplett.2008.02.025
27  Adiga, S. P., Curtiss, L. A. & Gruen, D. M. in *Nanodiamonds: Applications in Biology and Nanoscale Medicine* (ed Dean Ho) 35-54 (Springer US, 2010).
28  Los, J. H. & Fasolino, A. Intrinsic long-range bond-order potential for carbon: Performance in Monte Carlo simulations of graphitization. *Phys. Rev. B* **68**, 024107 (2003). https://doi.org/10.1103/PhysRevB.68.024107
29  Ganesh, P., Kent, P. R. C. & Mochalin, V. Formation, characterization, and dynamics of onion-like carbon structures for electrical energy storage from nanodiamonds using reactive force fields. *J. App. Phys.* **110**, 073506 (2011). https://doi.org/10.1063/1.3641984
30  Los, J. H., Pineau, N., Chevrot, G., Vignoles, G. & Leyssale, J.-M. Formation of multiwall fullerenes from nanodiamonds studied by atomistic simulations. *Phys. Rev. B* **80**, 155420 (2009). https://doi.org/10.1103/PhysRevB.80.155420
31  Dai, L. *et al.* Effect of micro-diamond and nano-polycrystalline diamond interfacial microstructure on OLC phase transition. *Chem. Eng. J.* **497**, 155613 (2024). https://doi.org/10.1016/j.cej.2024.155613
32  Dai, L. *et al.* Mechanism of phase transition from OLCs with different structures to nPCD at high temperature and high pressure. *J. Mater. Res. Technol.* **25**, 1322-1333 (2023). https://doi.org/10.1016/j.jmrt.2023.05.277
33  Powles, R. C., Marks, N. A. & Lau, D. W. M. Self-assembly of s${p}$-${2}$-bonded carbon nanostructures from amorphous precursors. *Phys. Rev. B* **79**, 075430 (2009). https://doi.org/10.1103/PhysRevB.79.075430
34  Chen, A. *et al.* Mechanism of graphitization from fragmental carbon to graphite film. *Mater. Today Commun.* **42**, 111529 (2025). https://doi.org/10.1016/j.mtcomm.2025.111529
35  Bidault, X. & Pineau, N. Dynamic formation of nanodiamond precursors from the decomposition of carbon suboxide (C3O2) under extreme conditions—A ReaxFF study. *J. Chem. Phys.* **149**, 114301 (2018). https://doi.org/10.1063/1.5028456

36  Enriquez, J. I. G. *et al.* Origin of the surface facet dependence in the thermal degradation of the diamond (111) and (100) surfaces in vacuum investigated by machine learning molecular dynamics simulations. *Carbon* **226**, 119223 (2024). https://doi.org/10.1016/j.carbon.2024.119223
37  Ostroumova, G., Orekhov, N. & Stegailov, V. Reactive molecular-dynamics study of onion-like carbon nanoparticle formation. *DRM* **94**, 14-20 (2019). https://doi.org/10.1016/j.diamond.2019.01.019
38  Orekhov, N., Ostroumova, G. & Stegailov, V. High temperature pure carbon nanoparticle formation: Validation of AIREBO and ReaxFF reactive molecular dynamics. *Carbon* **170**, 606-620 (2020). https://doi.org/10.1016/j.carbon.2020.08.009
39  Jones, A. P. *et al.* Structural characterization of natural diamond shocked to 60GPa; implications for Earth and planetary systems. *Lithos* **265**, 214-221 (2016). https://doi.org/10.1016/j.lithos.2016.09.023
40  Bogdanov, D. *et al.* Core growth of detonation nanodiamonds under high-pressure annealing. *RSC Adv.* **11**, 12961-12970 (2021). https://doi.org/10.1039/D1RA00270H
41  Thompson, A. P. *et al.* LAMMPS - a flexible simulation tool for particle-based materials modeling at the atomic, meso, and continuum scales. *Comput. Phys. Commun.* **271**, 108171 (2022). https://doi.org/10.1016/j.cpc.2021.108171
42  van Duin, A. C. T., Dasgupta, S., Lorant, F. & Goddard, W. A. ReaxFF: A Reactive Force Field for Hydrocarbons. *J. Phys. Chem. A* **105**, 9396-9409 (2001). https://doi.org/10.1021/jp004368u
43  Aktulga, H. M., Pandit, S. A., van Duin, A. C. T. & Grama, A. Y. Reactive Molecular Dynamics: Numerical Methods and Algorithmic Techniques. *SIAM J. Sci. Comput.* **34**, C1-C23 (2012). https://doi.org/10.1137/100808599
44  Jensen, B. D., Wise, K. E. & Odegard, G. M. Simulation of the Elastic and Ultimate Tensile Properties of Diamond, Graphene, Carbon Nanotubes, and Amorphous Carbon Using a Revised ReaxFF Parametrization. *J. Phys. Chem. A* **119**, 9710-9721 (2015). https://doi.org/10.1021/acs.jpca.5b05889
45  Li, K. *et al.* ReaxFF Molecular Dynamics Simulation for the Graphitization of Amorphous Carbon: A Parametric Study. *J. Chem. Theory Comput.* **14**, 2322-2331 (2018). https://doi.org/10.1021/acs.jctc.7b01296
46  de Tomas, C. *et al.* Transferability in interatomic potentials for carbon. *Carbon* **155**, 624-634 (2019). https://doi.org/10.1016/j.carbon.2019.07.074
47  Aghajamali, A. & Karton, A. Comparative Study of Carbon Force Fields for the Simulation of Carbon Onions. *Aust. J. Chem.* **74**, 709-714 (2021). https://doi.org/10.1071/CH21172
48  Damirchi, B. *et al.* ReaxFF Reactive Force Field Study of Polymerization of a Polymer Matrix in a Carbon Nanotube-Composite System. *J. Phys. Chem. C* **124**, 20488-20497 (2020). https://doi.org/10.1021/acs.jpcc.0c03509
49  Srinivasan, S. G., van Duin, A. C. T. & Ganesh, P. Development of a ReaxFF Potential for Carbon Condensed Phases and Its Application to the Thermal Fragmentation of a Large Fullerene. *J. Phys. Chem. A* **119**, 571-580 (2015). https://doi.org/10.1021/jp510274e
50  Yoon, K. *et al.* Atomistic-Scale Simulations of Defect Formation in Graphene under Noble Gas Ion Irradiation. *ACS Nano* **10**, 8376-8384 (2016). https://doi.org/10.1021/acsnano.6b03036
51  Jain, A. *et al.* Commentary: The Materials Project: A materials genome approach to accelerating materials innovation. *APL Materials* **1**, 011002 (2013). https://doi.org/10.1063/1.4812323
52  The Materials Project. Materials Data on C by Materials Project. (2020). https://doi.org/10.17188/1281384
53  X Xu, M. *et al.* Spontaneous formation of graphene-like stripes on high-index diamond C(331) surface. *Nanoscale Research Letters* **7**, 460 (2012). https://doi.org/10.1186/1556-276X-7-460
54  Ong, S. P. et al. Python Materials Genomics (pymatgen): A robust, open-source python library for materials analysis. *Comput. Mater. Sci.* **68**, 314-319 (2013). https://doi.org/10.1016/j.commatsci.2012.10.028

| | |
|---|---|
| 55 | Klotz, S., et al., Hydrostatic limits of 11 pressure transmitting media. *J. Phys. D: Appl. Phys.* **42(7)**, 07413, (2009). https://doi.org/10.1088/0022-3727/42/7/075413. |
| 56 | Shimizu, H., et al., High-Pressure Elastic Properties of Solid Argon to 70 GPa. *Phys. Rev. Lett.* **86(20)**, 4568-4571 (2001). https://link.aps.org/doi/10.1103/PhysRevLett.86.4568 |
| 57 | Tegeler, C., R. Span, and W. Wagner, A New Equation of State for Argon Covering the Fluid Region for Temperatures From the Melting Line to 700 K at Pressures up to 1000 MPa. *J. Phys Chem. Ref. Data*, **28(3)**, 779-850 (1999). https://doi.org/10.1063/1.556037 |
| 58 | Root, S., et al., Argon equation of state data to 1 TPa: Shock compression experiments and simulations. Phys. Rev. B 106(17), 174114 (2022). https://link.aps.org/doi/10.1103/PhysRevB.106.174114. |
| 59 | Hashemi, S.R., G. Barinovs, and G. Nyman, A ReaxFF molecular dynamics and RRKM ab initio based study on degradation of indene. *Frontiers in Astronomy and Space Sciences*, **10**, (2023). https://www.frontiersin.org/journals/astronomy-and-space-sciences/articles/10.3389/fspas.2023.1134729. |
| 60 | Bentley, J. L. Multidimensional binary search trees used for associative searching. *Commun. ACM* **18**, 509–517 (1975). https://doi.org/10.1145/361002.361007 |
| 61 | Maneewongvatana, S. & Mount, D. M. Analysis of approximate nearest neighbor searching with clustered point sets. *arXiv preprint cs/9901013* (1999). |
| 62 | Virtanen, P. *et al.* SciPy 1.0: fundamental algorithms for scientific computing in Python. *Nat. Methods* **17**, 261-272 (2020). https://doi.org/10.1038/s41592-019-0686-2 |
| 63 | Hagberg, A. A. et al. Exploring network structure, dynamics, and function using NetworkX. 7th Python in Science Conference (SciPy2008), 11-15 (2008). |
| 64 | Johnson, D. B. Finding All the Elementary Circuits of a Directed Graph. *SIAM Journal on Computing* **4**, 77-84 (1975). https://doi.org/10.1137/0204007 |
| 65 | Szwarcfiter, J. L. & Lauer, P. E. A search strategy for the elementary cycles of a directed graph. *BIT Numer. Math.* **16**, 192-204 (1976). https://doi.org/10.1007/BF01931370 |
| 66 | Loizou, G. & Thanisch, P. Enumerating the cycles of a digraph: A new preprocessing strategy. *Inf. Sci.* **27**, 163-182 (1982). https://doi.org/10.1016/0020-0255(82)90023-8 |
| 67 | Birmelé, E. *et al.* in *Proceedings of the twenty-fourth annual ACM-SIAM symposium on Discrete algorithms.* 1884-1896 (SIAM). |
| 68 | Gupta, A. & Suzumura, T. *Finding All Bounded-Length Simple Cycles in a Directed Graph*. A. Gupta, and T. Suzumura, arXiv preprint arXiv **2105**, 10094 (2021). |
| 69 | Coleman, S. P., Spearot, D. E. & Capolungo, L. Virtual diffraction analysis of Ni [0 1 0] symmetric tilt grain boundaries. *Model. Simul. Mater. Sci. Eng.* **21**, 055020 (2013). https://doi.org/10.1088/0965-0393/21/5/055020 |
| 70 | Bragg, W. H. & Bragg, W. L. The reflection of X-rays by crystals. *Proceedings of the Royal Society of London. Series A, Containing Papers of a Mathematical and Physical Character* **88**, 428-438 (1913). https://doi.org/doi:10.1098/rspa.1913.0040 |
| 71 | Pedregosa, F. *et al.* Scikit-learn: Machine Learning in Python. *J. Mach. Learn. Res.* **12**, 2825–2830 (2011). |
| 72 | Sapinski, M. Model of Carbon Wire Heating in Accelerator Beam. *CERN-AB-2008-030 BI. European Organization for Nuclear Research (CERN), Geneva, Switzerland*, 24 Jul 2008 (Sept 2008). |
| 73 | Stelmakh, S., Skrobas, K., Gierlotka, S., Vogel, S. C. & Palosz, B. Atomic structure and grain shape evolution of nanodiamond during annealing in oxidizing atmosphere from neutron diffraction and MD simulations. *DRM* **111**, 108177 (2021). https://doi.org/10.1016/j.diamond.2020.108177 |
| 74 | Amin, M. H., Mottalebizadeh, A. A. & Borji, S. Influence of cooling medium on detonation synthesis of ultradispersed diamond. *Diamond and Related Materials* **18**, 611-614 (2009). https://doi.org/10.1016/j.diamond.2008.10.048 |

75  Armstrong, M. R. *et al.* Highly ordered graphite (HOPG) to hexagonal diamond (lonsdaleite) phase transition observed on picosecond time scales using ultrafast x-ray diffraction. *J. Appl. Phys.* **132**, 055901 (2022). https://doi.org/10.1063/5.0085297
76  Blank, V. D., Kulnitskiy, B. A. & Nuzhdin, A. A. Lonsdaleite formation in process of reverse phase transition diamond–graphite. *DRM* **20**, 1315-1318 (2011). https://doi.org/10.1016/j.diamond.2011.08.009
77  Dolmatov, V. Y. The Influence of Detonation Synthesis Conditions on the Yield of Condensed Carbon and Detonation Nanodiamond Through the Example of Using TNT-RDX Explosive Mixture. *J. Superhard Mater.* **40**, 290-294 (2018). https://doi.org/10.3103/S1063457618040093
78  Heuser, B. *et al.* Release dynamics of nanodiamonds created by laser-driven shock-compression of polyethylene terephthalate. *Sci. Rep.* **14**, 12239 (2024). https://doi.org/10.1038/s41598-024-62367-7
79  Kuznetsov, V. L. & Butenko, Y. V. in *Synthesis, Properties and Applications of Ultrananocrystalline Diamond.* (eds Dieter M. Gruen, Olga A. Shenderova, & Alexander Ya Vul') 199-216 (Springer Netherlands).
80  Xu, Q. & Zhao, X. Bucky-diamond versus onion-like carbon: End of graphitization. *Phys. Rev. B* **86**, 155417 (2012). https://doi.org/10.1103/PhysRevB.86.155417
81  Zheng, Y. & Zhu, P. Carbon nano-onions: large-scale preparation, functionalization and their application as anode material for rechargeable lithium ion batteries. *RSC Adv.* **6**, 92285-92298 (2016). https://doi.org/10.1039/C6RA19060J
82  Zhang, C. *et al.* Synthesis of hollow carbon nano-onions and their use for electrochemical hydrogen storage. *Carbon* **50**, 3513-3521 (2012). https://doi.org/10.1016/j.carbon.2012.03.019
83  Alas, M. O., Alkas, F. B., Aktas Sukuroglu, A., Genc Alturk, R. & Battal, D. Fluorescent carbon dots are the new quantum dots: an overview of their potential in emerging technologies and nanosafety. *J. Mater. Sci.* **55**, 15074-15105 (2020). https://doi.org/10.1007/s10853-020-05054-y
84  Hui, S. Carbon dots (CDs): basics, recent potential biomedical applications, challenges, and future perspectives. *J. Nanopart. Res.* **25**, 68-112 (2023). https://doi.org/10.1007/s11051-023-05701-w
85  Zhang, C., Li, J., Zeng, X., Yuan, Z. & Zhao, N. Graphene quantum dots derived from hollow carbon nano-onions. *Nano Res.* **11**, 174-184 (2018). https://doi.org/10.1007/s12274-017-1617-0
86  Chua, C. K. *et al.* Synthesis of Strongly Fluorescent Graphene Quantum Dots by Cage-Opening Buckminsterfullerene. *ACS Nano* **9**, 2548-2555 (2015). https://doi.org/10.1021/nn505639q
87  Kuznetsov, V.L., et al., Theoretical study of the formation of closed curved graphite-like structures during annealing of diamond surface. J. Appl. Phys. 86(2), 863-870 (1999). https://doi.org/10.1063/1.370816
88  Luo, K., et al., Coherent interfaces govern direct transformation from graphite to diamond. Nature **607(7919)**, 486-491 (2022). https://doi.org/10.1038/s41586-022-04863-2
89  Zhang, H., Yan, Z., Zhang, H. & Chen, G. Graphitization of diamond: manifestations, mechanisms, influencing factors and functional applications. *Funct. Diamond* **5**, 2533896 (2025). https://doi.org/10.1080/26941112.2025.2533896
90  Li, Q., et al., Effect of stress state on graphitization behavior of diamond under high pressure and high temperature. DRM, **128**, 109241 (2022). https://www.sciencedirect.com/science/article/pii/S092596352200423X
91  Paul, S., K. Momeni, and V.I. Levitas, Shear-induced diamondization of multilayer graphene structures: A computational study. *Carbon* **167**, 140-147 (2020). https://www.sciencedirect.com/science/article/pii/S0008622320304796
92  Sakib, N., et al., From tilt-grained graphene to diamene: Exploring phase transformation and stability under extreme shear stress. DRM **155,** 112344 (2025). https://www.sciencedirect.com/science/article/pii/S0008622320304796

| | |
|---|---|
| 93 | Sakib, N., et al., Two-dimensional diamond-diamane from graphene precursor with tilt grain boundaries: Thermodynamics and kinetics. *DRM*, **145**, 111068 (2024). https://doi.org/10.1016/j.diamond.2024.111068 |
| 94 | Paul, S. and K. Momeni, Mechanochemistry of Stable Diamane and Atomically Thin Diamond Films Synthesis from Bi- and Multilayer Graphene: A Computational Study. *J. Phys. Chem. C* **123(25)**, 15751-15760 (2019). https://doi.org/10.1021/acs.jpcc.9b02149 |
| 95 | Ringstrand, B. S., Mogavero, B., Despard, J. T., Firestone, M. & Podlesak, D. Discontinuous density gradient fractionation of detonation soot for complete nanocarbon characterization. *AIP Conf. Proc.* **1979**, 100035 (2018). https://doi.org/10.1063/1.5044907 |
| 96 | Németh, P. *et al.* Diamond-Graphene Composite Nanostructures. *Nano Letters* **20**, 3611-3619 (2020). https://doi.org/10.1021/acs.nanolett.0c00556 |
| 97 | Németh, P. *et al.* Diaphite-structured nanodiamonds with six- and twelve-fold symmetries. *DRM* **119**, 108573 (2021). https://doi.org/10.1016/j.diamond.2021.108573 |
| 98 | Chen, G.-W. *et al.* The Transformation Mechanism of Graphite to Hexagonal Diamond under Shock Conditions. *JACS Au* **4**, 3413-3420 (2024). https://doi.org/10.1021/jacsau.4c00523 |